\documentclass[12pt]{article}
\usepackage{amssymb,amsmath} 
\usepackage{graphicx}
\usepackage{epsfig} 
\usepackage{latexsym}
\usepackage{psfrag}   
\usepackage{subfig}  
\usepackage{caption}
\usepackage{booktabs}  
\usepackage{braket}
\usepackage{mathtools}
\usepackage{protosem}
\usepackage{float}
\usepackage{changepage}
\usepackage{enumitem}
\usepackage{wasysym}
\usepackage[numbers,sort&compress]{natbib}
\usepackage{textcomp}
\usepackage[utf8]{inputenc}
\usepackage{pifont}
\usepackage{bbm} 
\usepackage[
      colorlinks=false,
      linkcolor=darkblue,  
      urlcolor=blue,    
      filecolor=blue,     
      citecolor=darkgreen, 
      linktocpage=true,
      pdfstartview=FitV,
      bookmarksopen=true    
      ]{hyperref}

\usepackage{tikz}

\usepackage{tikz-cd}
\usepackage{amsmath,amssymb,amsfonts,xspace} 
\usepackage{graphicx}
\usepackage[arrow,curve,matrix,arc]{xy}
\usepackage{epsfig,supertabular,longtable}

\usepackage{mathrsfs}
\usepackage[inner=2.9cm,outer=2cm]{geometry}
\usepackage[toc]{appendix}
\usepackage{color,soul} 
\usepackage{datetime}

\usepackage{setspace}

\usepackage{tikz}
\newcommand*\circled[1]{\tikz[baseline=(char.base)]{
            \node[shape=circle,draw,inner sep=2pt] (char) {#1};}}

\definecolor{darkblue}{rgb}{0.2, 0, 0.8}
\definecolor{darkgreen}{rgb}{0.2, 0.71, 0}

\numberwithin{equation}{section}

\newcommand{\Tr}{{\rm Tr}}

\newcommand{\bea}{\begin{eqnarray}}
\newcommand{\eea}{\end{eqnarray}}
\newcommand{\ba}{\begin{eqnarray}}
\newcommand{\ea}{\end{eqnarray}}

\newcommand{\beq}{\begin{equation}}
\newcommand{\eeq}{\end{equation} }
\newcommand{\beqa}{\begin{eqnarray}}
\newcommand{\eeqa}{\end{eqnarray}}
\newcommand{\beqar}{\begin{eqnarray*}}
\newcommand{\eeqar}{\end{eqnarray*}}

\definecolor{cardinal}{rgb}{0.6,0,0}
\definecolor{darkgreen}{rgb}{0,0.4,0}
\definecolor{purple}{rgb}{0.5, 0, 0.5}
\definecolor{golden}{rgb}{0.92, 0.7, 0}
\definecolor{midnight}{rgb}{0, 0, 0.5}
\definecolor{darkblue}{rgb}{0, 0, 0.7}

\begin{document}  
 \begin{flushright}
{\tt \small{IPhT-T18/119}} \\
\end{flushright}

\vspace*{1.0cm}

\phantom{AAA}
\vspace{-8.6mm}

\vspace{10mm}

\begin{center}

{\LARGE {\bf The full space of BPS multicenter states with pure D-brane charges}}

\bigskip

\vspace{15mm}

{\large
\textsc{ Pierre Heidmann$^{1}$}} and {\large
\textsc{ Swapnamay Mondal$^{2}$}}

\vspace{8mm}

$^1$Institut de Physique Th\'eorique,\\
Universit\'e Paris Saclay,\\
CEA, CNRS, F-91191 Gif sur Yvette, France \\
\bigskip
$^2$ International Centre for Theoretical Sciences, Tata Institute of Fundamental Research,\\
 Shivakote, Hesaraghatta, Bangalore 560089, India


\vspace{9mm} 
{\footnotesize\upshape\ttfamily pierre.heidmann @ ipht.fr, ~ swapno @lpthe.jussieu.fr } \\

\vspace{20mm}
 
\textsc{Abstract}

\end{center}

\begin{adjustwidth}{10mm}{10mm} 
 
\vspace{1mm}
\noindent
We investigate the space of BPS states in type IIA string theory on a T$^6 $  wrapped by one D6 brane and three D2 branes wrapping three disjoint 2-tori. This system of branes has 12 ground states. We show that these 12 states are all recovered as Coulomb branch BPS multicenter bound states, in which each center preserves 16 supercharges. Moreover, we show that these multicenter solutions can only exist with zero angular momentum, supporting the conjecture that all black hole microstates have zero angular momentum. For large charges, they might describe ``near-horizon limit" of fuzzballs.

\end{adjustwidth}

\thispagestyle{empty}
\newpage


\baselineskip=13pt
\parskip=3pt

\setcounter{tocdepth}{2}
\tableofcontents

\baselineskip=15pt
\parskip=3pt



\section{Introduction}

One of the greatest successes of string theory has been to reproduce microscopically the Bekenstein-Hawking entropy of several classes of supersymmetric black holes \cite{Strominger:1996sh}. This is supported by the idea that D-brane states describe the physics of black holes at low string coupling $g_s$. Investigating these two possible viewpoints: the microscopic D-brane picture and the macroscopic supergravity picture, and using quantities which are protected under change of $g_s$, such as the BPS index or the entropy, have brought a better understanding of the quantum structure of BPS black holes (see \cite{David:2002wn,Sen:2007qy} for reviews).

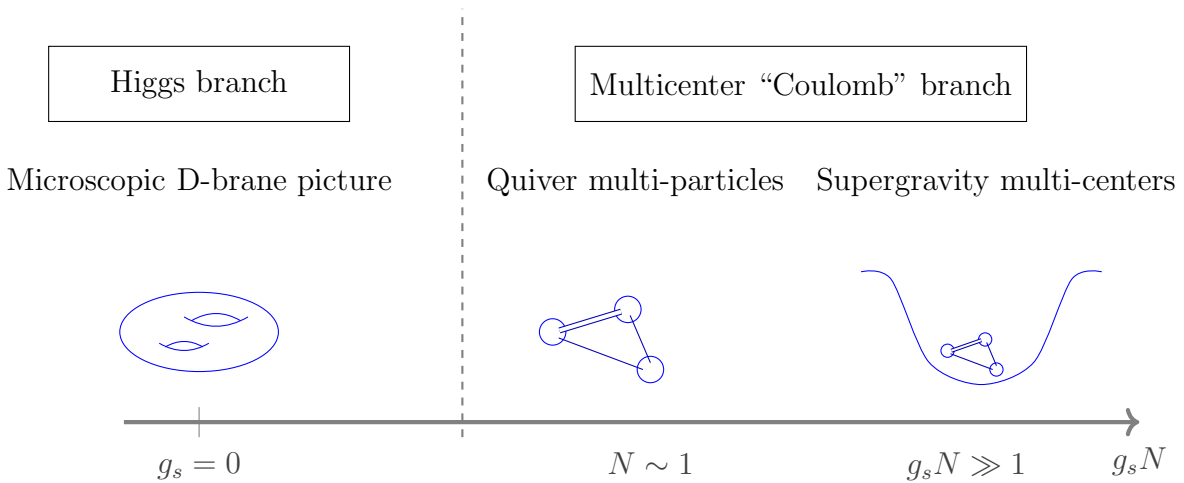
\begin{figure}
\begin{center}
\begin{tikzpicture}
\draw [->, ultra thick, gray](-8,0) -> (5.5,0) node[below=5,darkgray]{$g_s N$};
\draw[gray] (-7,-0.2)--(-7,0.2) node[below=13,darkgray]{$g_s =0$};
\draw (-1,0.2) node[below=13,darkgray]{$N \sim 1 $};
\draw (3.2,0.2) node[below=13,darkgray]{$g_s N \gg 1 $};
\draw[dashed,thick=2pt,gray] (-3.5,-0.2)--(-3.5,5.5);

 \draw[blue] (-7.45,1) to[bend left] (-6.95,1);
  \draw[blue] (-7.53,1.05) to[bend right]  (-6.87,1.05) ;
  
   \draw [blue] (-7.1,1.35) to[bend left] (-6.45,1.35);
  \draw[blue] (-7.2,1.4) to[bend right]  (-6.35,1.4) ;
  
  \draw[rotate=0,blue] (-7,1.2) ellipse (30pt and 15pt);
  
  \draw[rotate=0,blue] (-2.3,1.2) circle(5pt);
  \draw[rotate=0,blue] (-1.3,1.5) circle(5pt);
  \draw[rotate=0,blue] (-1,0.7) circle(5pt);
  \draw[double,double distance= 1.5pt,darkblue] (-2.21,1.22) -- (-1.39,1.47) ;
  \draw[darkblue]  (-2.22,1.12) -- (-1.1,0.7) ;
 \draw[darkblue]  (-1.27,1.42)  -- (-1,0.79) ;
 

\draw [blue] plot [smooth] coordinates {(1.8,2)(2.2,1.9) (2.7,0.8)(3.4,0.5)(4.1,0.8) (4.6,1.9)(5,2)};
(4.1,0.35)

  \draw[rotate=0,blue] (2.95,0.95) circle(2.5pt);
   \draw[rotate=0,blue] (3.6,0.7) circle(2.5pt);
 \draw[rotate=0,blue] (3.45,1.1) circle(2.5pt);
  \draw[double,double distance= 0.75pt,darkblue] (3,0.95) -- (3.4,1.09) ;
    \draw[darkblue]  (2.99,0.91) -- (3.55,0.7) ;
     \draw[darkblue]  (3.47,1.06)  -- (3.6,0.75) ;
   
 
 \draw (-9,4) rectangle (-5,5) node[pos=.5] {Higgs branch};
  \draw (-7,3.2) node {Microscopic D-brane picture};
  
\draw (-2,4) rectangle (4,5) node[pos=.5] {Multicenter ``Coulomb" branch};
\draw (-1.2,3.2) node {Quiver multi-particles};
    
\draw (3.6,3.2) node {Supergravity multi-centers};
\end{tikzpicture}
\caption{The description of a D-brane system in different regimes of parameters where $g_s$ is the string coupling and $N$ is the total number of branes.}
\label{fig:threeregimecoupling}
\end{center}
\end{figure}

Type II string theory on a six-dimensional Calabi-Yau manifold with D-branes wrapped on various cycles has been an extensively rich framework for state counting \cite{Dijkgraaf:1996xw,Maldacena:1999bp,Shih:2005qf,David:2006ud}. There are essentially three regimes in which one can work, depending on the value of $g_s$ and the number of D-branes $N$ (see Fig.\ref{fig:threeregimecoupling}): The Higgs branch is the one supporting the microscopic single D-brane picture of bound states and leads to exact results at infinitesimally small $g_s$. By increasing $g_s$, two scenarios are possible. The most admitted one is that the majority of the Higgs-branch states are recovered as single center black hole solutions. The second possible scenario is that the D-brane charges gather in several centers forming a molecule-like BPS bound states described by quiver quantum mechanics. We denote this branch as the multicenter ``Coulomb" branch\footnote{The Coulomb branch is not the same as the Coulomb branch obtained by moving the D2 and D6 branes away from each other. We are referring to the multicenter Coulomb branch where the separations between the centers cannot be modified freely.}. One can still define two regimes inside this branch: the supergravity regime, valid as long as $g_s N \gg 1$ where BPS states are well-described as macroscopic multicenter configurations, and the quiver regime, in which the system is described by a quiver quantum mechanics\footnote{See \cite{Denef:2002ru} for an exhaustive description of the different regimes.}.

In the present paper, we are interested in BPS solutions in the multicenter Coulomb branch of type IIA compactified\footnote{Some related investigation for Calabi Yau compactification has been performed in \cite{Gaddam:2016xum}} on T$^6$ wrapped by $Q_{D6}$ D6 branes and three D2 branes wrapping three disjoint 2-tori for $Q_{D6}=1$ or $2$.
In the Higgs branch, the number of states is 12 and 56 for $Q_{D6}=1$ and $Q_{D6}=2$ respectively \cite{Chowdhury:2014yca, Chowdhury:2015gbk,Shih:2005qf}. At a generic point of moduli space, the angular momentum of the microscopic D-brane states was found to be strictly zero. Motivated by this result obtained in the regime where $g_sN \ll 1$, and by the fact that in the regime where $g_sN \gg 1$ (when the black hole description is trustable) all microstates of a single centered supersymmetric black hole carry zero angular momentum, it was conjectured that microstates of such black holes continue to carry zero angular momentum at a generic point of the moduli space\footnote{Which includes $g_s$.}.

We test the conjecture in the multicenter Coulomb branch by building all the BPS multicenter bound states with the same D-brane charges and by investigating their possible moduli at infinity. Strictly speaking, we are in a multicenter Coulomb branch, far from the supergravity regime ($g_s N \ll 1$) and one should construct these configurations as solutions of quiver quantum mechanics. However, thanks to the work done in \cite{Denef:2000nb,Denef:2001xn,Denef:2001ix,Denef:2002ru,Bates:2003vx,Denef:2007vg}, one knows that the conditions of existence for quiver BPS multi-particle solutions are exactly the same as the ones for BPS multicenter solutions in the supergravity picture even if the supergravity solution is not reliable. 
As long as the geometry is not considered, the picture of charge vectors located at some centers still holds true for $g_s N \ll 1$. Thus, for a counting problem, one can use the supergravity framework which is more commonly understood to build the BPS solutions and study their general properties. 

A BPS multicenter solution is constructed by choosing a set of centers in a $\mathbb{R}^3$ base space which carry magnetic and electric charges corresponding to branes wrapping cycles of the transverse space and by choosing a set of moduli at infinity. Requiring the solutions to be supersymmetric and free of closed timelike curves restricts significantly their number. We consider only 16-supercharge centers following the intuition that black hole microstates are built with maximally-supersymmetric objects \cite{Bena:2013dka}. For $Q_{D6}=1$, we find exactly 12 BPS multicenter solutions. We show that they can't carry a non-zero angular momentum. This is in sharp contrast with the usual belief that it is easy to add momentum to multicenter solutions by slightly changing the moduli at infinity. From a supergravity point of view, it means that the possible asymptotics of the solutions are significantly restricted to the one preserving the zero angular momentum. In four dimensions, these solutions necessarily have AdS$_2\times$S$^2$  asymptotics. Furthermore, each solution can be dualized to a six-dimensional D1-D5-P frame where it can have an AdS$_3\times$S$^3$ asymptotics. However, the configurations we construct cannot give rise to asymptotically flat solutions in any frame.

The AdS$_2$ asymptotics of these geometries make them particularly interesting for the following reason. The near-horizon region of an extremal black hole develops an AdS$_2$ factor and this region decouples from the rest of the spacetime. This enables one to compute many physical black hole quantities, such as (quantum) entropy function \cite{Sen:2005wa,Sen:2007qy,Sen:2008vm},  solely in this near horizon region. Thus, an asymptotically AdS$_2$ geometry, carrying appropriate charges, can be thought of as replacing the horizon by some structures in the ``near horizon" region. It remains to be seen whether such ``near horizon" structures can be thought of as ``near horizon limit" of some full fledged fuzzball geometry, which lives in the same asymptotically flat ambient spacetime as the black hole. 
Despite this subtlety and the fact that we are really working in a regime where gravity can not be trusted, finding exactly $e^{S_{BH}}$ number of such microstate geometries, does inspire curiosity about the possibility that they indeed correspond to ``near-horizon limit" of some fuzzball geometries.

Moreover, it has been argued in \cite{Bena:2018bbd} that, in string theory, the ground states of the CFT$_1$ dual to AdS$_2$ must break conformal invariance. Thus, their bulk duals which are asymptotically AdS$_2$ geometries must break conformal invariance by having a scale in the IR. This argument eliminates the possibility that the single-center asymptotically AdS$_2$ solution (which has an infinite throat and preserves conformal invariance in the IR) is dual to any ground states of a non-trivial CFT$_1$. Our work confirms this expectation: From a microscopic counting the number of ground states of the CFT$_1$ corresponding to our D-brane system is 12. And indeed we have found exactly 12 asymptotically AdS$_2$ bulk multicenter solutions. Furthermore, all these solutions have a scale, given by the distances between the centers, and hence break conformal invariance in the IR, exactly as predicted by \cite{Bena:2018bbd}.

In section \ref{sec:MicrostateGeneralities}, we review the picture of BPS multicenter solutions as charge vectors $\Gamma_a$ localized at centers in a $\mathbb{R}^3$ base space and we review the constraints on the bound states we are looking for. In section \ref{sec:recapSen}, we review the state counting obtained in the Higgs branch at $g_s N \lll 1$ \cite{Chowdhury:2014yca, Chowdhury:2015gbk}. In section \ref{sec:sol(1,1,1,1)}, we explain the technical details of our own construction in the multicenter Coulomb branch. Finally, in section \ref{sec:discussion}, we discuss the main features of the 12 solutions, their description within the quiver-quantum-mechanics framework and the extension of our construction to D-brane bound states with $Q_{D6}=2$.


\section{BPS multicenter solutions}
\label{sec:MicrostateGeneralities}

We would like to construct states with four supercharges carrying D-brane charges for $g_s \ll 1$ and $N \sim 1$ in type IIA string theory on a T$^6$. In this regime of parameters, the states are described as multi-particle quiver bound states \cite{Denef:2002ru,Bates:2003vx,Denef:2007vg}. As explained in the introduction, we prefer to use supergravity tools. This does not compromise our analysis since we are interested in counting the number of states and in the charge profiles carried at each center. At this level, both the supergravity and quiver frames give the same results \cite{Denef:2000nb,Denef:2001ix,Denef:2007vg}.

\subsection{BPS multicenter solutions}

We work in the context of type IIA string theory on a $T^6=T^2 \times T^2\times T^2$ at the low energy limit $g_s\ll 1$. We consider multicenter BPS solutions where each center can carry D0-D2-D4-D6 charges wrapping the tori. They are stationary but generically non-static BPS bound states with four unbroken supersymmetries. They are specified by \emph{charge vectors} $\Gamma_a$ at each center and an \emph{asymptotic vector} $\Gamma_\infty$ incorporating the moduli at spatial infinity. We denote
\begin{alignat}{2}
&\Gamma_a &&= (Q_{D6},Q_{D4}^1,Q_{D4}^2,Q_{D4}^3;Q_{D2}^1,Q_{D2}^2,Q_{D2}^3,Q_{D0})_a, \nonumber\\
& \, && \equiv \left( q_a , k^1_a,k^2_a,k^3_a;l^1_a,l^2_a,l^3_a,m_a \right), \label{eq:chargevector} \\
&\Gamma_\infty&& \equiv \left( q_\infty , k^1_\infty,k^2_\infty,k^3_\infty;l^1_\infty,l^2_\infty,l^3_\infty,m_\infty \right). \nonumber
\end{alignat}
The solutions can also be expressed in terms of a set of 8 harmonic functions in $\mathbb{R}^3$ (we refer the reader to the Appendix \ref{app:BPSsolSUGRA} for a description of the field content and the resolution of the BPS equations.)
\begin{equation}
V=q_\infty+\sum_a \frac{q_a}{r_a} \, , \hspace{0.3cm}
K^I=k^I_\infty+\sum_a \frac{k^I_a}{r_a} \, , \hspace{0.3cm}
L_I=l^I_\infty+\sum_a \frac{l^I_a}{r_a} \, , \hspace{0.3cm}
M=m_\infty+\sum_a \frac{m_a}{r_a} \, ,
\label{eq:harmfunc}
\end{equation}

\noindent
where $I=1,2,3$ and $r_a$ is the three-dimensional distance from the a$^{th}$ \emph{center}. They can be incorporated to the charge and asymptotic vectors as
\begin{equation}
\Gamma \:\equiv\: (V,K^I, L_I, M) \:=\:\Gamma_\infty+ \sum_a \frac{\Gamma_a}{r_a}.
\end{equation}

\noindent
The ten-dimensional type IIA metric is given by
\begin{equation}
ds_{10}^2 \:=\: -\mathcal{I}_4^{\,-1/2} \, \left( dt \:+\: \omega \right) ^2 \:+\: \mathcal{I}_4^{\,1/2} \,ds(\mathbb{R}^3)^2 + \sum_{I=1}^3 \frac{\mathcal{I}_4^{\,1/2}}{\sqrt{Z_I V}} \, ds_I ^2,
\label{4Dmetric}
\end{equation}

\noindent
where \(ds_{I}^2\) is the internal metric of the $I^{th}$ 2-torus, the $Z_I$ are the warp factors giving rise to the three charges of the solution and $\mathcal{I}_4$ is the quartic invariant
\begin{equation}
\label{eq:I4}
\mathcal{I}_4 \equiv Z_1 Z_2 Z_3 V - \mu^2 V^2 .
\end{equation}

\noindent
As it has been argued previously, the geometry given by \eqref{4Dmetric} is not reliable in a regime where $N\sim 1$. In such a regime, the solutions should be seen as low-charge multi-particles which does not significantly backreact \eqref{4Dmetric}. However, the conditions of validity derived from the metric in the supergravity regime are still valid in the low-charge regime. Furthermore, from a holographic point of view, it is also interesting to have access to the geometrical features of the multicenter solutions when they are sent to the macroscopic regime\footnote{This can be achieved by simply multiplying the charge vectors by a constant $\Gamma ' = \Lambda \Gamma$ with $\Lambda \gg 1$.}.

To solve the BPS equations, the warp factors \(Z_I\) and the 1-form $\omega$ are constructed as follows,
\begin{eqnarray}
\nonumber
Z_I &=& L_I \:+\: \frac{1}{2}C_{IJK} \frac{K^J K^K}{V} \, , \\ \label{Z&kexpression}
\mu &=& \frac{1}{6} V^{-2} C_{IJK} K^I K^J K^K \:+\: \frac{1}{2} V^{-1} K^I L_I \:+\: \frac{M}{2} \, , \\ 
 \star_{(3)} d \omega &=&V \, d\mu \:-\: \mu \,dV \:-\: V \,Z_I \, d\left(\frac{K^I}{V} \right), \nonumber
\end{eqnarray}

\noindent
with $ C_{IJK} \:=\: \left| \epsilon_{IJK} \right|$.

\noindent
All BPS solutions need to be free of closed timelike curves and Dirac-Misner strings. The first condition requires the positivity of the \emph{quartic invariant} $\mathcal{I}_4$ (see Appendix \ref{app:BPSsolSUGRA}),

\begin{equation}
\label{eq:noctc}
\mathcal{I}_4  > 0\, .
\end{equation}

\noindent
 while the second restricts the inter-center distances $r_{ab}$ \cite{Denef:2000nb},
\begin{equation}
\label{eq:bubble1}
\sum_{b \neq a} \frac{ \langle \Gamma_a , \Gamma_b \rangle}{r_{ab}} = \langle\Gamma_\infty, \Gamma_a \rangle \,.
\end{equation}
\noindent
We have defined a symplectic product \(\langle\:,\:\rangle\) of 8-dimensional vectors $A=\left(A^0, A^I, A_I, A_0 \right)$ as
\begin{equation}
    \langle A,B \rangle \:\equiv\:  A^0 B_0-A_0 B^0 +A^I B_I-A_I B^I \, ,
    \label{symplprod}
\end{equation}  
\noindent
The equations \eqref{eq:bubble1}  are known as the \emph{bubble equations} and impose strong constraints on the space of solutions. The positivity of the quartic invariant is also a constraint which is difficult to manipulate. For our analysis, we will apply different versions which have been used in previous works. First, a necessary condition for the positivity of the quartic invariant is to satisfy the three inequalities
\begin{equation}
Z_I V\,\geq\,0 \,, \quad I=1,2,3\,, \qquad \mu \underset{r \rightarrow\infty}{\rightarrow} 0.
\label{eq:necCTCcondition}
\end{equation}
This condition is easier to derive and to check than \eqref{eq:noctc}. Moreover, in practice, it happens to be a sufficient condition. 
Second, we will apply a conjecture which reduces the condition regarding closed timelike curves to simpler algebraic conditions on the bubble equations (we refer the reader to the section 3 of \cite{Avila:2017pwi} for details).

Moreover, a generic BPS multicenter solution is stationary but not static. Its angular momentum, $\vec{J}$, is given by
\begin{equation}
\vec{J} \:\equiv\: \frac{1}{2} \sum_{a<b}  \langle \Gamma_a , \Gamma_b \rangle \, \hat{r}_{ab}\,,\qquad \hat{r}_{ab} \:\equiv\:\frac{\vec{r}_a - \vec{r}_b}{| \vec{r}_a - \vec{r}_b|}.
\label{eq:angMomentum}
\end{equation}
The value of the angular momentum is closely related to the moduli at infinity of the solution. Indeed, using the bubble equations one can show that
\ \begin{equation}
\vec{J} \:\equiv\:  \frac{1}{2}\sum_{a}  \langle\Gamma_\infty, \Gamma_a \rangle \, \vec{r}_a.
\label{eq:angMomentumModuli}
\end{equation}
Finally, being interested in specific D-brane charge configurations we express the quantized asymptotic D-brane charges according to the charges at each center \cite{Balasubramanian:2006gi}. They can be derived from flux-integrals of the RR gauge field forms $C^{(1)}$ and $C^{(3)}$ and their dual gauge fields $C^{(5)}$ and $C^{(7)}$ (see Appendix \ref{app:BPSsolSUGRA})
\begin{equation}
Q_{D6} \:=\:\sum_a q_a, \qquad Q_{D4}^I \:=\:\sum_a k_a^I, \qquad Q_{D2}^I \:=\: \sum_a l_a^I, \qquad Q_{D0} \:=\: \sum_a m_a.
\label{eq:branecharges}
\end{equation}
\subsection{Types of center}
\label{sec:centerspecies}

Since we are looking for BPS multicenter solutions with four supercharges, the choice of charge vectors \eqref{eq:chargevector} is restricted to the one which preserves supersymmetry. The maximally-supersymmetric centers are the two-charge supertubes and the Gibbons-Hawking (GH) centers. They preserve 16 supercharges and the two U(1) isometries of the three-dimensional base space. This is not an exhaustive list of possible string theory objects. One can also imagine objects as wiggly supertubes, four-dimensional superstratum and so on. They are less supersymmetric and they may break some of the U(1) isometries \cite{Bena:2015bea}. Following the Bena-Warner ansatz \cite{Bena:2013dka}, it is more likely that a system of low or pure D-brane charges \eqref{eq:branecharges} is fully or largely made of maximally-supersymmetric BPS objects.

\subsubsection{Two-charge supertube centers}

A two-charge supertube located at the $a^{th}$ center carries a dipole charge \(k_a\), two electric charges \(Q_a^{(b)}\) with $b\neq a$ and a momentum charge $m_a$ \cite{Mateos:2001qs}. In other words, a two-charge supertube of species $I$ has a D0 charge, a D4 charge $Q_{D4}^I$ and two D2 charges $Q_{D2}^J $ and $Q_{D2}^K$ where $I,$ $J$ and $K$ are all different and between 1 and 3. Consequently, there are three possible species of two-charge supertubes depending on which of the three possible D4 charges they can carry. If there is only one species of supertube in the center configuration, the solution can be dualized to a smooth spacetime in six dimensions. Otherwise, multi-supertube configurations are not smooth because each supertube sources different vector fields and one cannot render the geometry smooth using a particular vector as Kaluza-Klein vector, see \cite{Bena:2008wt}.

As an illustration, let us consider a two-charge supertube of species 1 located at the a$^{th}$ center with the charge vector $\Gamma_a$
\begin{equation}
\Gamma_a\:=\: (0,\,k_a,\,0,\,0 \: ;\:0,\,Q^{(2)}_a,\,Q^{(3)}_a,\,m_a).
\label{eq:Scenter}
\end{equation}
\noindent
In the analysis performed in \cite{Bena:2009en, Vasilakis:2011ki, Heidmann:2017cxt}, where it was derived that the condition to preserve 16 supercharges and the quantization of the charges require to fix the following parameters
\begin{equation}
m_a \:=\: \frac{Q_a^{(2)}Q_a^{(3)}}{k_a}\,, \qquad k_a,\,Q_a^{(2)},\,Q_a^{(3)},\,m_a  \in \mathbb{Z}.
\label{eq:constraintchargesS}
\end{equation}
In anticipation of what will follow, one can have supertube centers with some of its charges being zero. For instance, on can construct a center with
\begin{equation}
\Gamma_a\:=\: (0,\,0,\,0,\,0 \: ;\:0,\,0,\,Q_a,\,m_a),
\end{equation}
by simply imposing the D4 charge and one of the D2 charge of the supertube to be zero. We can similarly have
\begin{equation}
\Gamma_a\:=\: (0,\,k_a,\,0,\,0 \: ;\:0,\,0,\,Q_a,\,0).
\end{equation}
These types of centers should not be denoted as supertube centers since they are 16-supercharge simple D-brane centers. However, in our construction, we will abusively use the generic term ``supertube center" even for these objects.

\subsubsection{Gibbons-Hawking centers}

A Gibbons-Hawking center (GH) is a smooth center which carries D0-D2-D4-D6 charges. To be maximally supersymmetric, it has to satisfy the following additional regularity constraint 
\begin{equation}
\begin{split}
& l_b^I \:=\: -\frac{1}{2}C_{IJK} \frac{k_b^J k_b^K}{q_b} \, ,\qquad  m_a \:=\: \frac{1}{6}C_{IJK} \frac{k_b^I k_b^J k_b^K}{q_b^2} \,, \qquad q_a,\,k_a^{I},\,l_a^{I},\,m_a  \in \mathbb{Z}. .
\label{eq:constraintchargesGH}
\end{split}
\end{equation}
Thus, a GH center located at the a$^{th}$ center has a charge vector
\begin{equation}
\Gamma_a\:=\: \left(q_a,\,k_a^1,\,k_a^2,\,k_a^3 \: ;\:- \frac{k_a^2 k_a^3}{q_a},\,- \frac{k_a^1 k_a^3}{q_a},\,- \frac{k_a^1 k_a^2}{q_a},\,\frac{k_a^1 k_a^2 k_a^3}{q_a^2}\right).
\label{eq:GHcenter}
\end{equation}

The BPS multicenter solutions we are considering are bound-states of a certain number of such centers. They are sensible configurations when there are no Dirac-Misner strings between centers and no closed timelike curves. This is achieved imposing the bubble equations (\ref{eq:bubble1}), which fix the positions of the centers, and the global bound (\ref{eq:I4}). Provided those conditions are satisfied, we have a physical solution. However, experience shows that finding a set of appropriate parameters can involve a vast exploration.

\subsection{The moduli at infinity}
\label{sec:asymptotics}

The moduli at large distance of a multicenter BPS solution are encoded by the asymptotic vector $\Gamma_\infty$. From a macroscopic point of view, the constant terms in the harmonic functions given by $\Gamma_\infty$ fix the asymptotics of the solution, which can be directly seen from the behavior of the metric at large distance \eqref{4Dmetric}. For instance, it has been showed in \cite{Bena:2018bbd}, that having no constant terms in the eight harmonic functions ($\Gamma_\infty =0$) corresponds to asymptotically AdS$_2$ multicenter solutions. An asymptotically flat $\mathbb{R}^{4,1}$ multicenter solution can be obtained by having a constant term in each $L_I$ function and a constant term in $M$ ($\Gamma_\infty =(0,0,0,0;1,1,1,m_\infty)$ where $m_\infty$ is fixed to have $\sum_{a}  \langle\Gamma_\infty, \Gamma_a \rangle =0$ ). Last but not least, a multicenter solution is asymptotically AdS$_3$ in the dual six-dimensional D1-D5 frame when one $L_I$ has a constant term turning on ($\Gamma_\infty =(0,0,0,0;1,0,0,m_\infty)$ for instance) \cite{Bena:2011zw}.

In a regime far from the supergravity regime $N\sim1$, the moduli at infinity $\Gamma_\infty$ loses its geometrical meaning. However, one can still relate the microscopic multicenter solutions we are building to their macroscopic equivalents by sending $\Gamma \rightarrow \Lambda \Gamma$ with $\Lambda \gg 1$ where the geometry is trustable. So we can still relate our states to a certain type of asymptotics and then determine their holographic meaning.

The different choices of moduli at infinity affect drastically the existence of a multicenter solution because of the bubble equations \eqref{eq:bubble1} \cite{Heidmann:2017toappear}. Moreover, they affect significantly the value of the angular momentum \eqref{eq:angMomentumModuli}.

In this paper, we will first build multicenter solutions with zero angular momentum $\vec{J}$ following the results in \cite{Chowdhury:2014yca, Chowdhury:2015gbk}. According to \eqref{eq:angMomentumModuli}, one can impose $\langle\Gamma_\infty, \Gamma_a \rangle = 0$ for all centers. This does not necessarily imply that $\Gamma_\infty=0$ and that the solutions are asymptotically AdS$_2$. However, taking $\Gamma_\infty=0$ is always a possibility. Consequently, one can just consider that $\langle\Gamma_\infty, \Gamma_a \rangle = 0$ in all our computations and build the charge vectors afterwards. Then, nothing stops us to test what moduli at infinity are indeed compatible with those charges. In particular, one can try to impose some moduli at infinity which give an angular momentum to the solution. This will give a non-trivial test to the zero-angular momentum conjecture. What we will find will confirm the conjecture: no moduli at infinity which gives angular momentum to our states are compatible with the charges.

\section{Counting Higgs-branch states with pure D2 and D6 charges}
\label{sec:recapSen}

In this section, we give a brief account of the state counting performed in \cite{Chowdhury:2014yca,Chowdhury:2015gbk}. In these papers the authors considered a D-brane system with four unbroken supercharges in four dimensions, in a duality frame where all charges were Ramond-Ramond charges. This theory is obtained by compactification over a T$^6$ and has $32$ supercharges. Broken supersymmetries give rise to $32-4=28$ Goldstinos and 28 partner Goldstone bosons. The Goldstones and Goldstinos can be arranged in $7$ supersymmetric multiplets, in four dimensional terminology. Due to these Goldstino multiplets, the Witten index vanishes. In order to get something non-vanishing yet protected, one needs to consider the $14^{th}$ helicity supertrace \cite{Bachas:1996bp,Gregori:1997hi} 
\begin{align}
B_{14} &:=  - \frac{1}{14 !} \Tr{} \left[ (-1)^{2J_3} (2J_3)^{14} \right] \, .
\end{align}
This essentially removes the contribution from the Goldstino multiplets and computes Witten index in the rest of the theory. 

Microscopically, this D-brane system comprises 3 stacks of D2 branes, along three disjoint 2-cycles of the T$^6$, and 1 stack of D6 branes along T$^6$. In order to compute the index, it suffices to have the knowledge of only low energy dynamics of this configuration. This is described by a quantum mechanics living on the intersection point of the branes. Fields in this theory correspond to massless strings stretched between various branes. The spectrum can be arranged in 4 dimensional supersymmetric multiplets, with the understanding that they are $0+1$ dimensional fields.
For the rest of the section, we use four dimensional terminologies, with the understanding that everything is dimensionally reduced to $0+1$ dimension.

The Lagrangian, $L$, has the following schematic form
\begin{align}
L &= \sum_{i=1}^4 L^{(i)}_{\mathcal{N}=4} + \sum_{i,j=1 \atop i<j}^4 L^{(ij)}_{\mathcal{N}=2} + L_{\mathcal{N}=1} \, ,
\end{align}
where $i$ denotes the brane index. The first piece, $L^{(i)}_{\mathcal{N}=4}$, denotes the $\mathcal{N}=4$ super Yang-Mills theory living on the $i^{th}$ brane, preserving $16$ supercharges. Although for different stacks, these supercharges are different. The second piece, $L^{(ij)}_{\mathcal{N}=2}$, denotes the interaction of $i^{th}$ and $j^{th}$ brane, which preserves $8$ supercharges. These two pieces are determined by supersymmetry alone. Altogether these terms preserve only $4$ supercharges, since each of them preserve different supersymmetries. Supersymmetry allows for, but does not determine, other interactions preserving $4$ supercharges, denoted as $L_{\mathcal{N}=1}$. In \cite{Chowdhury:2014yca,Chowdhury:2015gbk}, the authors considered few terms of $L_{\mathcal{N}=1}$ and argued that higher order terms, although certainly present, do not affect the index computation.

Computing $B_{14}$, is same as computing the Witten index in the theory obtained by throwing away Goldstino multiplets, which are non-interacting. In \cite{Chowdhury:2014yca}, authors identified the Goldstones, which are bosonic counterparts of the Goldstinos, using physical reasoning. The problem now reduces to one of computing Witten index in supersymmetric quantum mechanics, which is known to be given by Euler number of the vacuum manifold. For small charges, this was computed in \cite{Chowdhury:2014yca,Chowdhury:2015gbk} and was found to be in agreement with existing computations of the same in other duality frames \cite{Shih:2005qf}.

Typically microscopic computations of the index are performed at a special point of moduli space, and in such cases the index receives contributions both from bosonic and fermionic states. The computations of \cite{Chowdhury:2014yca,Chowdhury:2015gbk} however required turning on some moduli and therefore going to a more generic point of moduli space. Consequently, all the ground states of the D-brane quantum mechanics were found to be bosonic, in particular they carried zero angular momentum\footnote{The connection with angular momentum is made by identifying Lefschetz $SU(2)$ of the vacuum manifold with the $SU(2)$ corresponding to rotations in $\mathbb{R}^3$ \cite{Denef:2002ru}.}. This feature also holds for the regime of moduli space, where the brane system is better described as a single-center\footnote{A single-center 1/8 BPS black holes in $\mathcal{N}=8$ theory can fragment in two half BPS black holes, if a certain inequality is satisfied by the charges carried by the black hole. The charge vectors considered in \cite{Chowdhury:2014yca,Chowdhury:2015gbk} were not of that kind.} supersymmetric black hole. Such black holes are also known to carry zero angular momentum. In view of these facts, the authors of \cite{Chowdhury:2015gbk} conjectured that at a generic point of moduli space, all microstates of a single-center BPS black hole carry zero angular momentum.

\section{Counting zero-momentum multicenter states with pure D2 and D6 charges}
\label{sec:sol(1,1,1,1)}

In this section, we investigate the same D-brane configuration in a different region of parameters. We focus only on the space of BPS multicenter bound states even if it is possible that some or even all the Higgs-branch states shift to different kinds of solutions as single-center solutions once $g_s$ is turned on. To start with, we do not expect that counting of the states in the multicenter Coulomb branch gives the same number as the counting in the Higgs branch. Our procedure is to build the BPS multicenter solutions first and then count them. Thus, we will be able to analyze directly the features of the solutions. 

We build zero-momentum multicenter solutions with the following set of asymptotic D-branes charges:
\begin{equation}
Q_{D6} \:=\: 1, \qquad Q_{D4}^I \:=\: 0, \qquad Q_{D2}^I \:=\: 1, \qquad Q_{D0} \:=\: 0 \, .
\label{eq:imposedbranecharges}
\end{equation}
With such low charges, the supergravity regime should not be valid and the geometry of the center depicted in the section \ref{sec:MicrostateGeneralities} should not be trustable. However, one can use the supergravity solutions to count states. In \cite{Denef:2000nb,Denef:2001xn,Denef:2002ru,Bates:2003vx,Denef:2007vg}, it has been shown that BPS multicenter solutions with low charges can still be depicted by charge vectors \eqref{eq:chargevector} satisfying the bubble equations \eqref{eq:bubble1} and a condition equivalent to the absence of closed timelike curves \eqref{eq:noctc}. Consequently, one can still use the supergravity framework detailed in section \ref{sec:MicrostateGeneralities} to build all the multicenter solutions satisfying \eqref{eq:imposedbranecharges}. 

Our approach consists in scanning analytically or numerically all the valid multicenter solutions formed by either GH or supertube centers starting with the family of three-center solutions then the family of four-center solutions and finally the five-center solutions\footnote{Two-center solutions with GH centers or supertube centers have 8 remaining supercharges. Hence they do not correspond to the system we study.}. The growing complexity of the analysis does not allow to scan solutions with more than five centers but we have a strong intuition that adding centers increases necessarily the global D-brane charges of the solution. Thus, if solutions exist, they should consist in few centers. The constraints which restrict the possible solutions are:
\begin{itemize}[noitemsep,topsep=1pt]
\setlength{\itemsep}{0.2\baselineskip}
\item The global D-brane charges \eqref{eq:imposedbranecharges}.
\item The bubble equations  \eqref{eq:bubble1}.
\item  The absence of closed timelike curves \eqref{eq:noctc}.
\item The constraints at each center depending on the nature of the center \eqref{eq:constraintchargesGH} or \eqref{eq:constraintchargesS}.
\end{itemize} 
We have found exactly 12 distinct three-center solutions satisfying all those constraints. This matches exactly the microscopic counting. Readers interested only in the main ideas of our analysis can skip the next subsections until section \ref{subsec:summary}.

\subsection{The family of three-center solutions}
\label{sec:3centeranalysis}

One needs at least one GH center in the configuration to have a non-zero D6 charge. We divide our analysis in three subfamilies:
\begin{itemize}[noitemsep,topsep=3pt]
\setlength{\itemsep}{0.2\baselineskip}
\item[-] Solutions with one GH center and two supertube centers.
\item[-] Solutions with two GH centers and one supertube.
\item[-] Solutions with three GH centers.
\end{itemize}
For the first subfamily, an analytic approach is possible. All the details of this analysis are given in Appendix \ref{app:anal2S1GH}. The main result is that there exist 12 inequivalent solutions in this subfamily.
As for the second and the third families, the number of parameters makes the analytic approach impossible. However, we have performed an efficient numerical analysis, fully detailed in Appendix \ref{app:num1S2GH}. We have scanned a significant part of the parameter space of the solutions by varying each GH charge $q_a$, $k_a^1$, $k_a^2$, $k_a^3$ and supertube charge $k_a$, $Q^{(b)}_a$ from -500 to 500. We didn't find any solution satisfying all the constraints in this domain of values.

In \cite{Bianchi:2017bxl}, the authors have tackled a similar issue by analyzing the parameter space of three-GH center solutions whose the total D6 charge is three and the three D2 charges are one. Interestingly, they have found that the total number of such multicenter solutions is also 12. This is half a coincidence with our computation. Even if the three-center solutions they study have a larger $Q_{D6}$\footnote{Total number of microscopic states with $Q_{D6}= 3$, $ Q_{D4}^I = 0$,  $Q_{D2}^I = 1$, $Q_{D0} =0$ is  actually 208 \cite{Shih:2005qf}.}, the form of their specific solutions are governed by the same type of permutations giving rise to the 12 states we found here. 

\subsection{The family of four-center solutions}

A four-center solution has more degrees of freedom than the previous solutions and the constraints are more complicated to deal with. This makes any analytic investigation very hard to perform. However, we have done a numerical analysis of the following subfamilies:
\begin{itemize}[noitemsep,topsep=3pt]
\setlength{\itemsep}{0.2\baselineskip}
\item[-] Solutions with one GH center and three supertube centers.
\item[-] Solutions with two GH centers and two supertube.
\item[-] Solutions with three GH centers and one supertube.
\item[-] Solutions with four GH centers.
\end{itemize}
For each subfamily, we have analyzed a significant part of the parameter space by varying all the parameters from -10 to 10. The details are given in the Appendix \ref{app:num4center}. Our final result is that there are no valid solutions with four centers.

\subsection{The family of five-center solutions}

For five-center solutions, even a scan of the parameter space is complicated. This is principally due to the number of parameters available and the complexity of the constraints. However, we have been able to pick randomly a huge number of solutions with the right global D-brane charges and check if they can satisfy the bubble equations and the absence of closed timelike curves at the same time. We did not find any. This gives good intuition that no solution with five centers exists as well.

\subsection{Summary}
\label{subsec:summary}

We have analyzed analytically and numerically a huge number of BPS multicenter solutions to find only 12 solutions satisfying all the constraints. This exactly matches the exact counting of the 12 Higgs-branch states \cite{Chowdhury:2014yca}. They are all recovered as Coulomb branch multicenter bound states. They belong to the family of three-center solutions with one GH center and two supertubes of different species. The 12 solutions are given in full detail in Table \ref{TableSol}. Moreover, as explained in section \ref{sec:centerspecies}, for most of the solutions found, the two-charge-supertube centers are actually simple fluxed D-brane centers\footnote{This means that some of D-brane charges of the two-charge supertube are zero.}. For instance, the six first solutions in Table \ref{TableSol} have a GH center and two D4-brane centers with an induced D2 charge. The six other solutions have one GH center, one two-charge-supertube center and one simple D2-brane center with an induced D0 charge. 

\usetikzlibrary{decorations}
\usetikzlibrary{decorations.pathmorphing}
\usetikzlibrary{decorations.pathreplacing}
\noindent
\begin{table} 
\caption{The 12 multicenter solutions with global D-brane charges $(Q_{D6},Q_{D4}^1,Q_{D4}^2,Q_{D4}^3;Q_{D2}^1,Q_{D2}^2,Q_{D2}^3,Q_{D0})=(1,0,0,0;1,1,1,0)$.}
\label{TableSol}
\begin{center}
\begin{tabular}{|c|l|c|}
  \hline
 N° & \phantom{\Bigg(} \begin{minipage}{0.62\textwidth} Charge vectors at each center \\ $\Gamma_a = (Q_{D6},Q_{D4}^1,Q_{D4}^2,Q_{D4}^3;Q_{D2}^1,Q_{D2}^2,Q_{D2}^3,Q_{D0})_a$ \end{minipage} & Center configuration  \\
  \hline
 	1 & \quad \parbox{0cm}{\begin{alignat}{2}
 		&\Gamma_0 &&\:=\:  (1,\,1,\,-1,\,0\:;\:0,\,0,\,1,\,0) \nonumber\\
 		&\Gamma_1 &&\:=\:  (0,\,-1,\,0,\,0\:;\:0,\,1,\,0,\,0)  \nonumber \\
 		&\Gamma_2 &&\:=\: (0,\,0,\,1,\,0\:;\:1,\,0,\,0,\,0) \nonumber
 	\end{alignat}}
 	&  \parbox{2.5cm}{\begin{tikzpicture}
\draw [gray](1,0) -- (3,0);
\filldraw [black] (1,0) circle (2pt) node[above=2] {1};
\filldraw [black] (2,0) circle (2pt) node[above=2] {0} ;
\filldraw [black] (3,0) circle (2pt) node[above=2] {2};
\draw[color=black,decorate,decoration={brace,raise=0.2cm},rotate=180]
(-3,0) -- (-2.1,0) node[below=8,pos=0.5] {r};
\draw[color=black,decorate,decoration={brace,raise=0.2cm},rotate=180]
(-1.9,0) -- (-1,0) node[below=8,pos=0.5] {r};
\end{tikzpicture}}
  \\
  \hline
   	2 &\quad \parbox{2cm}{\begin{alignat}{2}
 		&\Gamma_0 &&\:=\:  (1,\,-1,\,1,\,0\:;\:0,\,0,\,1,\,0) \nonumber\\
 		&\Gamma_1 &&\:=\:  (0,\,1,\,0,\,0\:;\:0,\,1,\,0,\,0)  \nonumber \\
 		&\Gamma_2 &&\:=\: (0,\,0,\,-1,\,0\:;\:1,\,0,\,0,\,0) \nonumber
 	\end{alignat}}
 	& \parbox{2.5cm}{\begin{tikzpicture}
\draw [gray](1,0) -- (3,0);
\filldraw [black] (1,0) circle (2pt) node[above=2] {1};
\filldraw [black] (2,0) circle (2pt) node[above=2] {0} ;
\filldraw [black] (3,0) circle (2pt) node[above=2] {2};
\end{tikzpicture}}
  \\
  \hline
   	3 &\quad \parbox{3cm}{\begin{alignat}{2}
 		&\Gamma_0 &&\:=\:  (1,\,0,\,1,\,-1\:;\:1,\,0,\,0,\,0) \nonumber\\
 		&\Gamma_1 &&\:=\:  (0,\,0,\,-1,\,0\:;\:0,\,0,\,1,\,0)  \nonumber \\
 		&\Gamma_2 &&\:=\: (0,\,0,\,0,\,1\:;\:0,\,1,\,0,\,0) \nonumber
 	\end{alignat}}
 	&  \parbox{2.5cm}{\begin{tikzpicture}
\draw [gray](0,0) -- (2,0);
\filldraw [black] (0,0) circle (2pt) node[above=2] {1};
\filldraw [black] (1,0) circle (2pt) node[above=2] {0} ;
\filldraw [black] (2,0) circle (2pt) node[above=2] {2};
\end{tikzpicture}}
  \\
  \hline
  	4 &\quad \parbox{2cm}{\begin{alignat}{2}
 		&\Gamma_0 &&\:=\:  (1,\,0,\,-1,\,1\:;\:1,\,0,\,0,\,0) \nonumber\\
 		&\Gamma_1 &&\:=\:  (0,\,0,\,1,\,0\:;\:0,\,0,\,1,\,0)  \nonumber \\
 		&\Gamma_2 &&\:=\: (0,\,0,\,0,\,-1\:;\:0,\,1,\,0,\,0) \nonumber
 	\end{alignat}}
 	&  \parbox{2.5cm}{\begin{tikzpicture}
\draw [gray](0,0) -- (2,0);
\filldraw [black] (0,0) circle (2pt) node[above=2] {1};
\filldraw [black] (1,0) circle (2pt) node[above=2] {0} ;
\filldraw [black] (2,0) circle (2pt) node[above=2] {2};
\end{tikzpicture}}
  \\
  \hline
  	5 &\quad \parbox{2cm}{\begin{alignat}{2}
 		&\Gamma_0 &&\:=\:  (1,\,1,\,0,\,-1\:;\:0,\,1,\,0,\,0) \nonumber\\
 		&\Gamma_1 &&\:=\:  (0,\,-1,\,0,\,0\:;\:0,\,0,\,1,\,0)  \nonumber \\
 		&\Gamma_2 &&\:=\: (0,\,0,\,0,\,1\:;\:1,\,0,\,0,\,0) \nonumber
 	\end{alignat}}
 	&  \parbox{2.5cm}{\begin{tikzpicture}
\draw [gray](0,0) -- (2,0);
\filldraw [black] (0,0) circle (2pt) node[above=2] {1};
\filldraw [black] (1,0) circle (2pt) node[above=2] {0} ;
\filldraw [black] (2,0) circle (2pt) node[above=2] {2};
\end{tikzpicture}}
  \\
  \hline
  	6 &\quad \parbox{2cm}{\begin{alignat}{2}
 		&\Gamma_0 &&\:=\:  (1,\,-1,\,0,\,1\:;\:0,\,1,\,0,\,0) \nonumber\\
 		&\Gamma_1 &&\:=\:  (0,\,1,\,0,\,0\:;\:0,\,0,\,1,\,0)  \nonumber \\
 		&\Gamma_2 &&\:=\: (0,\,0,\,0,\,-1\:;\:1,\,0,\,0,\,0) \nonumber
 	\end{alignat}}
 	&  \parbox{2.5cm}{\begin{tikzpicture}
\draw [gray](0,0) -- (2,0);
\filldraw [black] (0,0) circle (2pt) node[above=2] {1};
\filldraw [black] (1,0) circle (2pt) node[above=2] {0} ;
\filldraw [black] (2,0) circle (2pt) node[above=2] {2};
\end{tikzpicture}}
  \\
  \hline
  
\end{tabular}
\end{center}
\end{table}

\begin{table} 
\begin{center}
\begin{tabular}{|c|l|c|}
 \multicolumn{1}{c}{\phantom{N°}} &   \multicolumn{1}{c}{\phantom{\phantom{\Bigg(} \begin{minipage}{0.62\textwidth} Charge vectors at each center \\ $\Gamma_a = (Q_{D6},Q_{D4}^1,Q_{D4}^2,Q_{D4}^3;Q_{D2}^1,Q_{D2}^2,Q_{D2}^3,Q_{D0})$ \end{minipage} }} &  \multicolumn{1}{c}{\phantom{center configuration }} \\
  \hline
 	7 & \quad \parbox{0cm}{\begin{alignat}{2}
 		&\Gamma_0 &&\:=\:  (1,\,-1,\,0,\,0\:;\:0,\,0,\,0,\,0) \nonumber\\
 		&\Gamma_1 &&\:=\: (0,\,1,\,0,\,0\:;\:0,\,1,\,1,\,1) \nonumber \\
 		&\Gamma_2 &&\:=\:  (0,\,0,\,0,\,0\:;\:1,\,0,\,0,\,-1) \nonumber
 	\end{alignat}}
 	&  \parbox{2.5cm}{\begin{tikzpicture}
\draw [gray](1,0) -- (3,0);
\filldraw [black] (1,0) circle (2pt) node[above=2] {0};
\filldraw [black] (2,0) circle (2pt) node[above=2] {1} ;
\filldraw [black] (3,0) circle (2pt) node[above=2] {2};
\end{tikzpicture}}
  \\
  \hline
   	8 & \quad \parbox{0cm}{\begin{alignat}{2}
 		&\Gamma_0 &&\:=\:  (1,\,1,\,0,\,0\:;\:0,\,0,\,0,\,0) \nonumber\\
 		&\Gamma_1 &&\:=\: (0,\,-1,\,0,\,0\:;\:0,\,1,\,1,\,-1)  \nonumber \\
 		&\Gamma_2 &&\:=\: (0,\,0,\,0,\,0\:;\:1,\,0,\,0,\,1)  \nonumber
 	\end{alignat}}
 	&  \parbox{2.5cm}{\begin{tikzpicture}
\draw [gray](1,0) -- (3,0);
\filldraw [black] (1,0) circle (2pt) node[above=2] {0};
\filldraw [black] (2,0) circle (2pt) node[above=2] {1} ;
\filldraw [black] (3,0) circle (2pt) node[above=2] {2};
\end{tikzpicture}}
  \\
  \hline
   	9 & \quad \parbox{0cm}{\begin{alignat}{2}
 		&\Gamma_0 &&\:=\:  (1,\,0,\,-1,\,0\:;\:0,\,0,\,0,\,0) \nonumber\\
 		&\Gamma_1 &&\:=\: (0,\,0,\,1,\,0\:;\:1,\,0,\,1,\,1) \nonumber \\
 		&\Gamma_2 &&\:=\:  (0,\,0,\,0,\,0\:;\:0,\,1,\,0,\,-1) \nonumber
 	\end{alignat}}
 	&  \parbox{2.5cm}{\begin{tikzpicture}
\draw [gray](1,0) -- (3,0);
\filldraw [black] (1,0) circle (2pt) node[above=2] {0};
\filldraw [black] (2,0) circle (2pt) node[above=2] {1} ;
\filldraw [black] (3,0) circle (2pt) node[above=2] {2};
\end{tikzpicture}}
  \\
  \hline
   	10 & \quad \parbox{0cm}{\begin{alignat}{2}
 		&\Gamma_0 &&\:=\:  (1,\,0,\,1,\,0\:;\:0,\,0,\,0,\,0) \nonumber\\
 		&\Gamma_1 &&\:=\: (0,\,0,\,-1,\,0\:;\:1,\,0,\,1,\,-1) \nonumber \\
 		&\Gamma_2 &&\:=\:  (0,\,0,\,0,\,0\:;\:0,\,1,\,0,\,1) \nonumber
 	\end{alignat}}
 	&  \parbox{2.5cm}{\begin{tikzpicture}
\draw [gray](1,0) -- (3,0);
\filldraw [black] (1,0) circle (2pt) node[above=2] {0};
\filldraw [black] (2,0) circle (2pt) node[above=2] {1} ;
\filldraw [black] (3,0) circle (2pt) node[above=2] {2};
\end{tikzpicture}}
  \\
  \hline
 11 & \quad \parbox{0cm}{\begin{alignat}{2}
 		&\Gamma_0 &&\:=\:  (1,\,0,\,0,\,-1\:;\:0,\,0,\,0,\,0) \nonumber\\
 		&\Gamma_1 &&\:=\: (0,\,0,\,0,\,1\:;\:1,\,1,\,0,\,1) \nonumber \\
 		&\Gamma_2 &&\:=\:  (0,\,0,\,0,\,0\:;\:0,\,0,\,1,\,-1) \nonumber
 	\end{alignat}}
 	&  \parbox{2.5cm}{\begin{tikzpicture}
\draw [gray](1,0) -- (3,0);
\filldraw [black] (1,0) circle (2pt) node[above=2] {0};
\filldraw [black] (2,0) circle (2pt) node[above=2] {1} ;
\filldraw [black] (3,0) circle (2pt) node[above=2] {2};
\end{tikzpicture}}
  \\
  \hline
   	12 & \quad \parbox{0cm}{\begin{alignat}{2}
 		&\Gamma_0 &&\:=\:  (1,\,0,\,0,\,1\:;\:0,\,0,\,0,\,0) \nonumber\\
 		&\Gamma_1 &&\:=\: (0,\,0,\,0,\,-1\:;\:1,\,1,\,0,\,-1) \nonumber \\
 		&\Gamma_2 &&\:=\:  (0,\,0,\,0,\,0\:;\:0,\,0,\,1,\,1) \nonumber
 	\end{alignat}}
 	&  \parbox{2.5cm}{\begin{tikzpicture}
\draw [gray](1,0) -- (3,0);
\filldraw [black] (1,0) circle (2pt) node[above=2] {0};
\filldraw [black] (2,0) circle (2pt) node[above=2] {1} ;
\filldraw [black] (3,0) circle (2pt) node[above=2] {2};
\end{tikzpicture}}
  \\
  \hline
\end{tabular}
\end{center}
\end{table}

We do not have indisputable arguments that having more centers will not give rise to other valid solutions but only good intuition. Usually, adding centers increases the global D-brane charges. Another difficulty in adding centers follows from the fact that these centers must carry negative D-brane charges, in order to keep the total D-brane charges intact. However usually centers with negative D-brane charges are tricky, when it comes to the $Z_I\,V \geq 0$, i.e. absence of closed timelike curves. For these reasons, we can consider our analysis exhaustive even if we have analyzed configuration with few centers. 

\section{Discussion}
\label{sec:discussion}

\subsection{Features of the twelve solutions}
\label{subsec:FeatureofSol}


The 12 solutions found are all BPS three-center solutions formed by one GH center and two other 16-supercharge centers. Although we looked for all possible center configurations, it happens that the regular solutions we found have their centers lying on a line and hence are all axisymmetric. The fact that the index is reproduced by counting configurations with collinear centers was also observed in \cite{Manschot:2010qz} and, given the very complicated algebra that required our physical solutions to be collinear, we do not believe this is a coincidence. 

The center configuration has also a rescaling symmetry $r_{ab} \rightarrow \lambda \, r_{ab}$. This is a consequence of having zero angular momentum which implies that the bubble equations have no right-hand terms \eqref{eq:bubble1} and so a scaling degree of freedom. When the solutions are asymptotically AdS$_2$, this scaling symmetry arises from the conformal invariance of the dual theory (the only scale in the system is the one introduced by the breaking of the conformal symmetry in the IR). When the solutions are asymptotically AdS$_3$, then the solutions with different values of the inter-center distances are physically inequivalent (the solutions have two scales, one given by the distances between the centers and one given by the constant in the harmonic function\footnote{In a macroscopic supergravity picture, the second scale is the one where the AdS$_2$ very-near-horizon region begins.}). 

We can now investigate what kind of moduli at infinity determined by $\Gamma_\infty$ are compatible with our multicenter solutions. The fact that microstates of black holes have necessarily zero angular momentum at any point of the moduli space has been conjectured in \cite{Sen:2009vz, Dabholkar:2010rm, Chowdhury:2015gbk}. A single-center black hole solution clearly has this property. This is because when defining the near horizon AdS$_2$ path integral, that is needed to compute quantum entropy function, one needs to fix the charges. And angular momentum appears as a charge in this path integral. Nevertheless, one can also argue that a single-center black hole does not correspond to any pure state of the CFT$_1$ dual to AdS$_2$ \cite{Bena:2018bbd}, and the zero-angular momentum assymptotically AdS$_2$ solutions dual to pure states of the CFT$_1$ will have a non-trivial angular momentum when embedded in asymptotically AdS$_3$ geometries. This is what happens in the solutions constructed in \cite{Bena:2018bbd}. \\
We would like to understand whether our 12 multicenter solutions can also develop a non trivial angular momentum when embedded in an asymptotically AdS$_3$ space, or whether they are incompatible with $\vec{J} \neq 0$.

\begin{itemize}
\item There are moduli at infinity which preserve $\vec{J}=0$. As it has been discussed in section \ref{sec:asymptotics}, this implies that $\langle\Gamma_\infty, \Gamma_a \rangle = 0$ for all the centers. Due to the numerous zeroes in the charge vectors, there are many possible $\Gamma_\infty$. However, one needs to carefully check that these $\Gamma_\infty$-s do not induce closed timelike curves. We give the list of possible moduli at infinity compatible with our solutions. Because all the 12 solutions have similar properties, it suffices to list the possibilities corresponding to the first solution in the Table \ref{TableSol}. 
We list only the interesting moduli at infinity and their holographic meaning:
\begin{itemize}
\item[-] $\Gamma_\infty =0$. As depicted in section \ref{sec:asymptotics}, this corresponds to asymptotically AdS$_2$ microstate geometries. The 12 bound states are thus holographically dual to the 12 ground states of a CFT$_1$. In \cite{Bena:2018bbd}, it has been argued that the ``$e^S$" ground states of a non-topological CFT$_1$ must break conformal invariance. So, all their bulk duals must have a scale, as it is actually the case for multicenter or superstratum solutions\footnote{For multicenter solutions, the scale is determined by the inter-center distances.}. A single-center asymptotically AdS$_2$ solution does not have any scale and thus should not be dual to any ground state of a non-trivial CFT$_1$. Our results corroborate this conjecture. 
\item[-] $\Gamma_\infty =(0,0,0,0;0,0,1,0)$. The meaning of this choice can be seen only from a six-dimensional point of view in the dual D1-D5-P frame. In this frame, the solution turns out to be an asymptotically AdS$_3$ state. 
\end{itemize} 
Both kinds of moduli at infinity impose $\vec{J}=\vec{0}$ and do not have any impact on the center configuration.
\item There are also moduli at infinity which do not preserve the symplectic products $\langle\Gamma_\infty, \Gamma_a \rangle$ and therefore give rise to a finite angular momentum, $\vec{J}$, without affecting the D-brane charges. This type of moduli requires to solve once again the bubble equations \eqref{eq:bubble1}. Because the inter-center distances can be as small as possible, one might expect that a change of $\langle\Gamma_\infty, \Gamma_a \rangle$ can be absorbed by a very small change of distances in $\sum_b \frac{ \langle \Gamma_a , \Gamma_b \rangle}{r_{ab} +\delta r_{ab}}$. This is unfortunately not guaranteed especially for axisymmetric multicenter configuration\footnote{This particular issue will be treated in general in an upcoming paper \cite{Heidmann:2017toappear}.}. Our multicenter solutions illustrate this feature. One can prove that any moduli at infinity which has $\langle\Gamma_\infty, \Gamma_a \rangle\neq 0$ and which does not induce closed timelike curves is incompatible with our solutions. Thus all our solutions are incompatible with having a non-zero angular momentum\footnote{One might expect some twist in the story in the special case when angular momentum is $1/2$. This is because upon quantization, a classical configuration with angular momentum $J_3$, is known to have angular momentum $J_3-1/2$. This implies classical configurations with angular momentum $1/2$ would quantum mechanically carry vanishing angular momentum. We do not explore this curious case in this paper though.
}.
\end{itemize}
This corroborates in a different regime of $g_S$ the conjecture that every microstate of a single centered supersymmetric black hole must have zero angular momentum at any generic point of moduli space \cite{Sen:2009vz, Dabholkar:2010rm, Chowdhury:2015gbk}. However, from an holographic point of view, the solutions can be either asymptotically AdS$_3$ \footnote{When dualized to the D1-D5-P duaity frame in type IIB.} or asymptotically AdS$_2$ solutions. 

For the particular choice of moduli at infinity $\Gamma_\infty =(0,0,0,0;0,0,1,0)$, only 6 of the 12 solutions survive. Those are the solutions with $K^3=0$, i.e. the solutions labelled by 1, 2, 7, 8, 9 and 10 in the Table \ref{TableSol}.  The six others have unsolvable bubble equations. At this particular point on the moduli space, one can wonder where the other 6 solutions are to make the total number of solutions to be 12. Two scenarios are possible. The first one is that at this point of the moduli space, there are only 6 multicenter solutions and the 6 others are of a different nature. The second one is that 6 brand new multicenter solutions appear which reversely cannot be sent to the point of moduli space where $\Gamma_\infty =0$. This interesting issue will be treated in future works.

\subsection{Quantum Effects}
Thus far, our analysis has been classical. However the charges considered being small, one might expect quantum effects to significantly modify our analysis. Of particular interest is the fate of angular momentum when quantum mechanical effects are taken into account. A classical configuration with angular momentum $J_3$ is known to furnish a spin $J_3-1/2$ representation of $SO(3)$, when quantum effects are taken into account \cite{Denef:2002ru}. This makes the present case, which corresponds to $J_3=0$, particularly puzzling. Since there is nothing called a spin $-1/2$ representation, either the angular momentum does not receive quantum correction in this particular case, or there are no supersymmetric ground states corresponding to the multicenter configurations carrying classically zero angular momentum.

To settle this question, one must analyse these multicenter configurations quantum mechanically. The framework for this has been laid down in \cite{Denef:2002ru}, where it has been shown that such multicenter configurations (equivalently intersecting D-branes wrapping various cycles of a Calabi-Yau threefold) are described by $\mathcal{N}=4$ quiver quantum mechanics. For an exhaustive discussion, we refer the interested reader to \cite{Douglas:1996sw,Denef:2002ru,Denef:2007vg,Bena:2012hf,Dabholkar:2012nd}.

Briefly, field content of a quiver quantum mechanics is encoded in a quiver diagram, which has as many nodes as centers and as arrows between nodes. The $a^\text{th}$ node corresponds to a vector multiplet with $U(N_a)$ gauge symmetry, with $N_a$ being determined by the charge vector $\Gamma_a$. For primitive $\Gamma_a$, which is the case at hand, one has $N_a=1$. Thus we shall restrict to Abelian quivers. 
For $\langle \Gamma_{a}, \Gamma_b \rangle > 0$, there are $\langle \Gamma_{a}, \Gamma_b \rangle$ arrows stretching from $a^{th}$ node to $b^{th}$ node, each corresponding to a hypermultiplets in the $U(1)\times\overline{U(1)}$ bifundamental. The dynamics of the fields is captured by a quiver Lagrangian fully determined by the charge vectors, their intersection products, Fayet-Iliopoulos parameters (henceforth referred to as FI parameters) and superpotential (when the quiver has loops).

Each of the 12 solutions corresponds to a three node Abelian quiver. For each quiver, we define the unique triplet of integer intersection product $(a,b,c)\equiv(\langle \Gamma_{i}, \Gamma_j \rangle,\langle \Gamma_{j}, \Gamma_k \rangle , \langle \Gamma_{k}, \Gamma_i \rangle )$ where $i,j,k$ are three different integers between 0 and 2 in order to satisfy $a \geq b >0$ and $c>0$. Each quiver has a closed loop and ($a,b,c$) satisfies the three triangle inequalities, $a+b\geq c$  and permutations. Three-nodes quivers have been extensively studied in the literature, particularly the ones with a closed loop \cite{Denef:2002ru,Denef:2007vg,Bena:2012hf,Dabholkar:2012nd}, for non-zero FI parameters. However we have vanishing FI parameters, which makes a lot of difference. In the following we analyze the relevant quiver.

All 12 solutions correspond to $(a,b,c)=(2,1,1)$ or some permutations thereof. Thus, the quiver under discussion is the following
\begin{equation}
\begin{aligned}
\begin{tikzcd}
&\circled{1} \ar[dr,"Y"]& \\
\circled{1} \ar[ur,"X"] & & \circled{1} \ar[ll,"\text{$C_1, C_2$}"]   \\
\end{tikzcd} 
\end{aligned} \, .
\end{equation}
The D-term equations read as follows
\begin{align}
|X|^2 - |C_1|^2 - |C_2|^2 &= 0 \nonumber \\
|Y|^2 - |X|^2 &= 0 \\
- |Y|^2 + |C_1|^2 + |C_2|^2 &= 0 \nonumber\, .
\end{align}
Following \cite{Bena:2012hf}, we assume a generic cubic superpotential
\begin{align}
W &= w_i XY C_i \, ,
\end{align}
which gives the following F-term equations
\begin{align*}
w_i Y C_i &= 0, \qquad w_i X C_i = 0,\qquad  XY =0 \, .
\end{align*}
The last equation requires either $X$ or $Y$ to vanish. But the second D-term equation implies that both $X$ and $Y$ vanishes. The remaining D-term equations imply $C_i=0$. Thus, the moduli space is a single point and hence furnishes a spin 0 representation of Lefschetz $SU(2)$. Thus we indeed have one quantum ground state with vanishing angular momentum. It is interesting to note that the vacuum preserves $U(1) \times U(1)$. 

In order to decide whether this should be counted as pure-Higgs state or not, it is instructive to briefly describe the similar computation carried out in \cite{Bena:2012hf}, but with non-zero FI parameters. After one of the three variables has been set to zero, in order to satisfy F-term equations, the D-term equations define a product of two projective spaces and the remaining F-term equation define a complete intersection manifold in this product of projective spaces. The Betti numbers of the complete intersection manifold, can be read out from those of the ambient space, except middle cohomology, where there can be extra states called ``pure-Higgs states".

In the present case, the FI parameters are zero and then the projective space collapses to a point. Consequently, the cohomology of the ambient space consists of a single state, which lies in the middle cohomology. So it is not very clear whether to count this state as pure-Higgs or not. We have however showed that each of the 12 quivers has only one ground state which hopefully leave the total of 12 states as expected from \cite{Chowdhury:2014yca, Chowdhury:2015gbk}.

The argument for general $(a,b,c)$ is not very different as discussed in Appendix \ref{abcapp}.

\subsection{The solutions with $Q_{D6}=2$}

As it has been done in \cite{Chowdhury:2015gbk}, one can extend our construction to compute the number of BPS states in supergravity framework with global D-brane charges:
\begin{equation}
Q_{D6} \:=\: 2, \qquad Q_{D4}^I \:=\: 0, \qquad Q_{D2}^I \:=\: 1, \qquad Q_{D0} \:=\: 0
\label{eq:imposedbranecharges2}
\end{equation}
The index is known to be 56 for these charges \cite{Shih:2005qf,Chowdhury:2015gbk}. We have been able to find only 18 BPS bound states formed by three centers similar to the ones described in section \ref{sec:sol(1,1,1,1)} and 12 BPS unbound states formed by four axisymmetric centers. The special aspect of those 12 unbound states is that one of the centers does not interact with the other centers and can be placed anywhere on the axis of the center configuration. We did not consider these solutions because they are similar to the ``real Coulomb" branch where the centers can move freely. Since we did not count the unbound Coulomb-branch configuration with non-interacting D6 center and three D2 centers, we are not counting these unbound states either. The charge vectors of the 18 three-center bound states are given in Table \ref{TableSol2} in the Appendix \ref{app:QD6=2}.

One can also describe the 18 states as three-node quivers with a closed loop and vanishing FI parameters. The only difference is that for $Q_{D6}=2$, 9 solutions are given by the triplet ($a,b,c$)=(3,2,1) and the 9 others give ($a,b,c$)=(3,1,2) (see Table \ref{TableSol2}). This does not affect our general argument discussed in Appendix \ref{abcapp}. Each quiver has only one ground state which keeps the total of 18 states.

\noindent
Our present construction technique does not allow easily to go beyond the multicenter solutions with GH centers or supertube centers and to find the 38 missing states. This will require more work in future projects. However, one can already have an idea of where these states may come from:
\begin{itemize}
\item Adding extra gauge fields and preserving the U(1)$\times$U(1) isometry of the base space. Following \cite{Gauntlett:2002nw,Bena:2013ora}, one can add for example a fourth massless abelian gauge field to the configuration, which corresponds to changing the fluxes on the T$^6$. The solutions will be slightly more complex but they will remain U(1)$\times$U(1) invariant and one may hope, along the line of \cite{Manschot:2010qz}, that these will contribute to the index.
\item Constructing configurations which break the U(1)$\times$U(1) isometry.  Such objects may include wiggly supertubes \cite{Bena:2010gg} or  superstratum configurations \cite{Bena:2015bea,Bena:2016ypk}. In four dimensions, these solutions give rise to KK modes along the two U(1) fibers of the base space and do not correspond to supergravity solutions.
\end{itemize}

\section{Conclusion}

In this paper, we have investigated the space of states of the following D-brane configuration
\begin{equation}
(Q_{D6},Q_{D4}^1,Q_{D4}^2,Q_{D4}^3\,;\,Q_{D2}^1,Q_{D2}^2,Q_{D2}^3,Q_{D0}) \:=\:(Q_{D6},0,0,0;1,1,1,0),\: \text{ with } Q_{D6}=1 \text{ or } 2.
\end{equation}
We have reviewed a computation in the Higgs branch where the states are depicted as microscopic single D-brane bound states \cite{Chowdhury:2014yca, Chowdhury:2015gbk}. The number of states is 12 and 56 for $Q_{D6}=1$ and $Q_{D6}=2$ respectively. These states have been argued to carry zero angular momentum.

We have tackled the same issues in the multicenter Coulomb branch. Multicenter BPS bound states are characterized by specifying charge vectors carried by center-like particles in the three-dimensional base space and a moduli at infinity \cite{Denef:2002ru}. All the multicenter BPS bound states can be retrieved from the microscopic D-brane bound states when $g_s \rightarrow 0$ but the opposite is not necessarily true. 

\noindent
For $Q_{D6}=1$, we have found exactly 12 BPS multicenter bound states. We have confirmed the counting from a quiver description by showing that the corresponding 12 quivers have only one supersymmetric ground state each. Furthermore these ground states carry zero angular momentum. This is in exception with common wisdom that quantum mechanically the angular momentum of such configurations shifts by $1/2$. The exception is made possible by vanishing of FI parameters. In this instance, all the microscopic D-brane states are recovered as BPS multicenter bound states and no single-center state should exist. This conveys the idea that BPS multicenter microstates which are types of fuzzballs in the macroscopic regime do not correspond to a peculiar part of the overall space of states of a certain D-brane system. Furthermore, we have shown that the 12 multicenter solutions carry necessarily zero angular momentum at this point of the moduli space giving greater weight to the zero angular momentum conjecture. From a supergravity point of view, this means that they are incompatible with having flat asymptotics. 

\noindent
For $Q_{D6}=2$, only 18 BPS multicenter bound states have been found. Using quiver quantum mechanics, we have shown in the Appendix \ref{abcapp} that they correspond to 18 supersymmetric ground states, each carrying zero angular momentum. We expect more multicenter solutions to exist. Indeed, our construction essentially focuses on U(1)$\times$U(1) invariant centers carrying 16-supercharges. One can expect, for configurations with more than pure D-brane charges, that less-isometric solutions exist. Such centers may be more exotic and less supersymmetric, such as wiggly supertubes. This investigation will lead to future projects. Nevertheless, the 18 solutions found also confirm the zero angular momentum conjecture.


\section*{Acknowledgments}

We are grateful to Iosif Bena for useful advice and to Boris Pioline for helpful discussions. The work of PH was supported by a CDSN grant from ENS Lyon and by the ANR grant Black-dS-String  ANR-16-CE31-0004-01. The work of SM was supported partly by CEFIPRA grant 5204-4 and most part of it was conducted within the ILP LABEX (ANR-10-LABX-63) supported by French state funds managed by ANR within the Investissements d'Avenir program (ANR-11-IDEX-0004-02).


\appendix
\section{More about BPS multicenter solutions in type IIA string theory}

\label{app:BPSsolSUGRA}

The solutions we consider in the following are supersymmetric configurations carrying various D0, D2, D4 and D6 charges in type IIA string theory. We assume a compactification by employing $T^6=(T^2)^3$ as the internal space. Our setup of D-brane charges \eqref{eq:branecharges} has exactly 3 massless abelian vector multiplets which corresponds to the STU model.

The action of the STU model is completely determined by the constant symmetric tensor $C_{IJK}$ which is equal to $\vert \varepsilon_{IJK} \vert$ for a diagonal $T^6$. Wrapped D-branes are incarnated in charged point centers in the four-dimensional base space which source the vector multiplet electrically (D0 and D2 charges) or magnetically (D4 and D6 charges). We are interested in solutions with two U(1) isometries. The solutions are uniquely determined by a complex structure moduli of $T^6$ and by an eight-dimensional vector $\Gamma$ of harmonic 3-forms (usually we omit the 3-form feature of the vector by considering the (D0,~D2,~D4,~D6)-basis of 3-forms and by writing $\Gamma$ as a eight-dimensional scalar vector in this basis). We denote
\begin{equation}
\Gamma \:\equiv\: (V,K^I, L_I, M) \:=\: \Gamma_\infty + \sum_a \frac{\Gamma_a}{r_a}.
\end{equation}
\noindent
The eight harmonic functions $\{V,K^I, L_I, M\}$ have poles at each center labeled by $a$, with charges given by the \emph{charge vectors} $\Gamma_a\equiv \left( q_a , k^1_a,k^2_a,k^3_a;l^1_a,l^2_a,l^3_a,m_a \right)$ with integer components and an asymptotic behavior given by the \emph{asymptotic vector} $\Gamma_\infty$ depending on the moduli at spacial infinity. 

All the timelike supersymmetric field configurations with two U(1) isometries  have a metric
\begin{equation}
ds_{4}^2 \:=\: -\mathcal{I}_4^{\,-1/2} \, \left( dt \:+\: \omega \right) ^2 \:+\: \mathcal{I}_4^{\,1/2} \,ds_{\mathbb{R}^3}^2,
\label{4Dmetricapp}
\end{equation}
\noindent
where $Z_I$ are the warp factors giving rise to the three charges of the solution and $\mathcal{I}_4$ is the quartic invariant
\begin{equation}
\label{eq:I4app}
\mathcal{I}_4 \equiv Z_1 Z_2 Z_3 V - \mu^2 V^2 .
\end{equation}
\noindent 
From the metric \eqref{4Dmetricapp}, the absence of closed timelike curves is straightforwardly guaranteed when the quartic invariant is positive $\mathcal{I}_4>0$.

The dilaton $\Phi$, the Neveu-Schwarz (NSNS) potential $B^{(2)}$, the Ramond-Ramond (RR) potentials $C^{(1)}$ and $C^{(3)}$ are given by \cite{Balasubramanian:2006gi,DallAgata:2010srl}
\begin{alignat}{2}
& e^{-2\Phi} && ~=~ \frac{V^3 Z_1 Z_2 Z_3}{\mathcal{I}_4^{\,3/2}}\, , \nonumber \\
& B^{(2)} && ~=~ \sum_{I=1}^3 B_I^{(2)} \, dT_I = \sum_{I=1}^3 \left(\frac{K^I}{V} - \frac{\mu}{Z_I}\right)\, dT_I \,, \nonumber \\
& C^{(1)} && ~=~ A - \frac{\mu \, V^2}{\mathcal{I}_4} (dt +\omega) \,, \\
& C^{(3)} && ~=~ \sum_{I=1}^3 C_I^{(3)} \wedge dT_I = \sum_{I=1}^3 \left[- \frac{dt+ \omega}{Z_I} +\left(\frac{K^I}{V} - \frac{\mu}{Z_I}\right) A + w^I \right] \wedge dT_I \,, \nonumber
\end{alignat}
where $dT_I$ is the volume form on the I$^{th}$ 2-torus and $A$ is a KK one-form satisfying $\star_{(3)} dA = dV$. One can derive from those expressions the NSNS 3-forms $H^{(3)}$, the RR field strengths $F^{(2)}$ and $F^{(4)}$ and their dual gauge fields $(C^{(5)},F^{(6)})$ and $(C^{(7)},F^{(8)})$ (see section 2.2 of \cite{DallAgata:2010srl} for more details). These field configurations become a solution when the following set of \emph{BPS equations}, defined on the three-dimensional base space, is satisfied 
\begin{eqnarray}
d\star_{(3)}d \,Z_I &~=~& \frac{C_{IJK}}{2} \,d\star_{(3)}d\, \left(\frac{K_J K_K}{V} \right)\, , \label{eq:BPSeq1}\\
\star_{(3)}d \,w_I &~=~& - \,d K_I \, , \label{eq:BPSeq2} \\
\star_{(3)}d \,\omega &~=~& V\,d\mu -\mu\,dV - V Z_I \,d\left(\frac{K_I}{V}\right) \,.  \label{eq:BPSeq3}
\end{eqnarray}
The equations \eqref{eq:BPSeq1} and \eqref{eq:BPSeq3} give
\begin{eqnarray}
\nonumber
Z_I &=& L_I \:+\: \frac{1}{2}C_{IJK} \frac{K^J K^K}{V} \, , \\ \label{Z&kexpressionapp}
\mu &=& \frac{1}{6} V^{-2} C_{IJK} K^I K^J K^K \:+\: \frac{1}{2} V^{-1} K^I L_I \:+\: \frac{M}{2} \,,
\end{eqnarray}
and the integrability condition of equation \eqref{eq:BPSeq3} yields the \emph{bubble equations}
\begin{equation}
\label{eq:bubbleapp}
\sum_{ b\neq a} \frac{ \langle \Gamma_a , \Gamma_b \rangle}{r_{ab}} = \langle\Gamma_\infty , \Gamma_a \rangle \, ,
\end{equation}
where the symplectic product has been defined in \eqref{symplprod}.

\noindent
Finally, one can compute the overall D-brane charges of the solution defined as integrals of the RR field strengths over cycles of the form $S_\infty^2 \times T_I^2$ where $S_\infty^2$ is the asymptotic two-sphere of the three-dimensional base space:
\begin{equation}
\begin{aligned}
&Q_{D6} \:=\: \int_{S_\infty^2} dC^{(1)} \:=\:\sum_a q_a, \\
& Q_{D4}^I \:=\: \int_{S_\infty^2 \times T_I^2} dC^{(3)}\:=\: \sum_a k_a^I, \\
&Q_{D2}^I \:=\: C_{IJK} \int_{S_\infty^2 \times T_J^2 \times T_K^2} dC^{(5)}\:=\: \sum_a l_a^I, \\
& Q_{D0} \:=\: \int_{S_\infty^2 \times T^6} dC^{(7)} \:=\: \sum_a m_a.
\label{eq:branechargesapp}
\end{aligned}
\end{equation}


\section{Analysis of three-center solutions}
\label{app:anal3center}
\subsection{Analytic investigation of solutions with two supertube and one GH centers}
\label{app:anal2S1GH}

We review our method to construct zero angular momentum BPS multicenter solutions  with global D0 and D4 charges being 0 and D6 and D2 charges being 1 starting from solutions with two supertube centers of different species and one GH center.
\begin{itemize}
\item We start with the full parameter space of solutions. The GH center is the $0^{th}$ center with charges $q$, $\kappa_1$, $\kappa_2$ and $\kappa_3$. We consider also a two-charge supertube of species 1 located at the $1^{st}$ center with charges $k_1$, $Q^{(2)}_1$ and $Q^{(3)}_1$ and a two-charge supertube of species 2 located at the $2^{nd}$ center with charges $k_2$, $Q^{(1)}_2$ and $Q^{(3)}_2$. The general form of the eight harmonic function is:
\begin{alignat}{3}
V &= \frac{q}{r_0} \, , \qquad && M &&= \frac{\kappa_1 \kappa_2 \kappa_3}{q^2 \,r_0}+\frac{Q_1^{(2)}Q_1^{(3)}}{k_1 \, r_1}+\frac{Q_2^{(1)}Q_2^{(3)}}{k_2\,  r_2} \,, \nonumber\\
K^{1} &= \frac{\kappa_1}{r_0}+ \frac{k_1}{r_{1}} \, , \qquad && L_{1} &&=-\frac{\kappa_2 \kappa_3}{q \,r_0}+ \frac{Q^{(1)}_2}{r_2} \, ,\nonumber\\
K^{2} &= \frac{\kappa_2}{r_0}+\frac{k_2}{r_{2}} \, , \qquad && L_{2} &&=-\frac{\kappa_1 \kappa_3}{q \,r_0}+\frac{Q^{(2)}_1}{r_1} \, ,\label{2supharmonicfunction}\\
K^{3} &= \frac{\kappa_3}{r_0} \, , \qquad && L_{3} &&=  -\frac{\kappa_1 \kappa_2 }{q \,r_0}+ \frac{Q^{(3)}_1}{r_1} +\frac{Q^{(3)}_2}{r_2} \, .\nonumber
\end{alignat}
First, we want to impose the values of the eight global D-brane charges. This reduces the number of free parameters to two. One can express everything in terms of $k_1$ and $k_2$:
\begin{alignat}{1}
\{q,\kappa_1,\kappa_2,\kappa_3\} &\:=\: \{1,- k_1,-k_2,0\} \nonumber \\
\{Q^{(2)}_1,Q^{(3)}_1\} &\:=\: \{1,\frac{k_1 (1 + k_1 k_2)}{k_1 - k_2}\} \label{eq:2paramfamily} \\
\{Q^{(1)}_2,Q^{(3)}_2\} &\:=\: \{1,\frac{k_2 (1 + k_1 k_2)}{k_2 - k_1}\} \nonumber 
\end{alignat}
\item For zero angular momentum three-center solutions, the bubble equations take the following simple form
\begin{equation}
\begin{split}
\label{eq:bubble2}
\frac{\Gamma_{01} }{r_{01}} + \frac{\Gamma_{02} }{r_{02}} = 0 \, , \\
-\frac{\Gamma_{01} }{r_{01}} + \frac{\Gamma_{12} }{r_{12}} = 0 \, , \\
-\frac{\Gamma_{02} }{r_{02}} - \frac{\Gamma_{12} }{r_{12}} = 0 \, ,
\end{split}
\end{equation}
which are easily solved by
\begin{equation}
r_{02} \:=\: - \frac{\Gamma_{02}}{\Gamma_{01}}\, r_{01} \, , \qquad r_{12} \:=\:  \frac{\Gamma_{12}}{\Gamma_{01}}\, r_{01} \,.
\end{equation}
We have used the notation $\Gamma_{IJ}= \langle \Gamma_{I},\Gamma_{J} \rangle $.
We notice that the solutions are invariant under rescaling of inter-center distances $r_{IJ} \rightarrow \lambda \, r_{IJ}$. That is why, $r_{01}$ remains a free parameter all along the construction. \\
Furthermore, the solution corresponds to a physical center configuration if and only if it satisfies the triangle inequality
\begin{equation}
\begin{split}
&\left( r_{01} + r_{02} - r_{12}    \right) \left( r_{01} - r_{02} + r_{12}   \right) \left( -r_{01} + r_{02} + r_{12} \right) \:\geq\:0 \\ 
&\iff \left( 1 - \frac{\Gamma_{02}}{\Gamma_{01}}  - \frac{\Gamma_{12}}{\Gamma_{01}}    \right) \left( 1 + \frac{\Gamma_{02}}{\Gamma_{01}}  + \frac{\Gamma_{12}}{\Gamma_{01}}    \right) \left( -1 - \frac{\Gamma_{02}}{\Gamma_{01}}  + \frac{\Gamma_{12}}{\Gamma_{01}}   \right) \:\geq\:0
\label{triangleinequ}
\end{split}
\end{equation} 
which constrains the two-dimensional parameter space of $k_1$ and $k_2$  significantly.
\item The solution must have a positive quartic invariant \eqref{eq:I4} to guarantee the absence of closed timelike curve. One can either use the conjecture in \cite{Avila:2017pwi} or the condition \eqref{eq:necCTCcondition}. Small number of centers makes the second option to be the simplest. We expand $Z_I V$ around each center. We find that $Z_I V\,\geq\,0$ imposes
\begin{equation}
\begin{split}
&q \,Q^{(1)}_2 \,\geq\, 0\, , \qquad q\, Q^{(2)}_1 \,\geq\, 0\, ,\qquad  \frac{q\,Q^{(3)}_1-k_1 k_2}{r_{01}}+\frac{q\,Q^{(3)}_2-k_1 k_2}{r_{02}} \,\geq\, 0 \, ,\\
 \qquad &q\, Q^{(3)}_1 +  k_1 k_2 \left(\frac{r_{01}}{r_{12}}-1\right) \,\geq\, 0\, , \qquad q\, Q^{(3)}_2 + k_1 k_2 \left( \frac{r_{02}}{r_{12}} -1\right) \,\geq\, 0\, ,
 \label{noCTC}
\end{split}
\end{equation} 
which further constrains the parameter space defined by $k_1$ and $k_2$. We remind the reader that these conditions are not necessarily sufficient to be free of closed timelike curves. One needs to check once those conditions satisfied that the quartic invariant is indeed positive.
\item Last but not least, one has to impose all the charges in the harmonic functions \eqref{2supharmonicfunction} to be integer.
\end{itemize}

\noindent
After few simplifications, the equations \eqref{triangleinequ} and \eqref{noCTC} are   satisfied if $k_1$ and $k_2$ satisfy
\begin{equation}
(k_1 \,>\, 0 \, \text{  and  } -\frac{1}{k_1} \,\leq\, k_2 \,\leq\, 0) \text{  or  } (k_1 \,<\, 0 \, \text{  and  } 0 \,\leq\, k_2 \,\leq\, -\frac{1}{k_1}) \text{  or  } (k_1 \,=\, 0 \, \text{  and  } |k_2| \,\geq \,1).
\label{eq:paramspacedomain}
\end{equation} 
Requiring each charge of the harmonic functions to be integer restricts \eqref{eq:paramspacedomain} to six possible values $(k_1,k_2) = \{(0,1),\,(0,-1),(1,0),\,(1,-1),\,(-1,0),\,(-1,1)\}$. \\
We can repeat exactly the same procedure with solutions of two supertubes of species 1 and 3 and solutions of two supertubes of species 2 and 3. By carefully counting the redundancies, we have a final count of 12 inequivalent solutions. Their charge vectors as well as their center configuration are given in detail in Table \ref{TableSol}. Their main and common features are that the center configurations are axisymmetric with a $U(1)$ symmetry and all centers carry D-brane charges of value -1, 0 or 1. Moreover, as explained in section \ref{sec:centerspecies}, for most of the solutions found, the two-charge-supertube centers are actually fluxed D-brane centers. The six first solutions in Table \ref{TableSol} have a GH center and two D4-brane centers with an induced D2 charge. The six other solutions have one GH center, one two-charge-supertube center and one simple D2-brane center with an induced D0 charge.  \\
We have carefully checked that the quartic invariant is strictly positive for all solutions found and that they are not related by gauge transformations. \\
One can also wonder why we do not consider configurations with two supertube centers of the same species. This is straightforward to check that such configurations are strictly incompatible with the global D-brane charges we impose \eqref{eq:imposedbranecharges}.

\subsection{Numerical analysis of solutions with one supertube and two GH centers and solutions with three GH centers}
\label{app:num1S2GH}

We review our numerical method which shows that there exists no valid solutions satisfying \eqref{eq:imposedbranecharges} with one supertube and two GH centers or with three GH centers. The number of parameters of such solutions makes an analytic approach difficult. The steps of our numerical analysis were the following:
\begin{itemize}
\item First, we start with the most general solutions. The solutions with one supertube and two GH centers form a family of 11 parameters whereas the solutions with three GH centers form a family of 12 parameters. 
\item We fix 8 parameters by imposing the global D-brane charges \eqref{eq:imposedbranecharges}. 
\item We run the other free parameters from -500 to 500. Each value corresponds to one particular solution. For each one, firstly we check if the solution has integer charges, secondly if the solutions of the bubble equations can give physical center configurations \eqref{triangleinequ}, and thirdly if the quartic invariant is positive. Checking the positivity of the quartic invariant is the hardest part. We have principally used the conjecture postulated in \cite{Avila:2017pwi}. This conjecture drastically simplifies the loop computations. It allows to check the positivity of the quartic invariant all over the $\mathbb{R}^3$ base space by checking an algebraic condition on a matrix derived from the bubble equations. This conjecture should work for multicenter solutions with GH centers only. However, one can mathematically consider supertube center as a limit of a GH center. For instance, one can obtain \eqref{eq:Scenter} from \eqref{eq:GHcenter} by taking the limit $\epsilon \rightarrow 0$ with
\begin{equation}
q_a = -\epsilon\, k_1 \, , \quad k_a^1 = k_1 \, , \quad k_a^2= \epsilon \,Q_a^{(3)} \, , \quad k_a^3= \epsilon \,Q_a^{(2)}.
\end{equation}
Thus, we can extend the conjecture to our solutions.
\end{itemize}

We did not find any solutions satisfying all the conditions in the huge range of parameters we have scanned. Furthermore, from the previous section we have a good intuition that if a solution exists the charges should be small. Consequently, one can say that our numerical analysis suggests that there is no solution of three GH centers or one supertube and two GH centers satisfying \eqref{eq:imposedbranecharges}.

\section{Analysis of four-center and five-center solutions}
\label{app:num4/5center}

\subsection{Analysis of four-center solutions}
\label{app:num4center}

We perform a similar analysis as in section \ref{app:num1S2GH}. The main goal is to scan a significant part of the parameter space looking for BPS four-center solutions satisfying \eqref{eq:imposedbranecharges}.
\begin{itemize}
\item As before, we start with the most general solutions. The solutions with three supertube and one GH centers form a family of 13 parameters, the solutions with two supertube and two GH centers form a family of 14 parameters, the solutions with one supertube and three GH centers form a family of 15 parameters and the solutions with four GH centers form a family of 16 parameters, . 
\item We fix 8 parameters by imposing the global D-brane charges \eqref{eq:imposedbranecharges}. 
\item We run the remaining parameters from -5 to 5 (the range of values is smaller than in section \ref{app:num1S2GH} due to the higher number of free parameters). For each value, we check if the solution is a valid BPS multicenter solution:
\begin{itemize}
\item[-] First, we check if all the harmonic-function charges are integer.
\item[-] Second, we check if the solution of the bubble equation can give rise to a physical center configuration. Because we have four centers and not three, this step is more complex than the one in the previous section. Indeed, we have to check four triangle inequalities as \eqref{triangleinequ} for each face of the tetrahedron formed by the four centers plus an angle inequality at one vertex of the tetrahedron.
\item[-] Third, we check the absence of closed timelike curves as in the previous section.
\end{itemize} 
\end{itemize}
No solution have been found in the range of values. One can realistically extend this result to all four-center BPS solutions.

\subsection{Analysis of five-center solutions}
\label{app:num5center}

The number of parameters and the complexity of the constraints for five-center configurations make the numerical scan of the parameter space impossible. However, we have randomly generated some solutions and checked if they are valid and physical. The main idea is to fix as many parameters as possible using the equations (the global D-brane charges, the bubble equations) and pick random values for the other parameters and check if they satisfy all the inequations (the absence of closed timelike curves, the triangle inequalities etc...). We have generated a significant number ($\sim 10^3$) of five-center solutions focusing on solutions with low charges at the centers , we find no valid solutions. This tends to argue that no five-center solutions with pure D6 and D2 charges exist.

\section{Configurations with $Q_{D6}=2$}
\label{app:QD6=2}

In this section, we give to the interested reader the charge vectors of the 18 three-center solutions with one GH center and two 16-supercharge centers with global D-brane charges $(Q_{D6},Q_{D4}^1,Q_{D4}^2,Q_{D4}^3\,;\,Q_{D2}^1,Q_{D2}^2,Q_{D2}^3,Q_{D0}) \:=\:(2,0,0,0;1,1,1,0)$. They are given in the Table \ref{TableSol2}. The center configurations are axisymmetric and look like
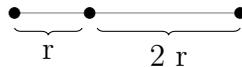
\begin{figure}[h]
\begin{center}
\begin{tikzpicture}
\draw [gray](1,0) -- (4,0);
\filldraw [black] (1,0) circle (2pt) node[above=2] {};
\filldraw [black] (2,0) circle (2pt) node[above=2] {} ;
\filldraw [black] (4,0) circle (2pt) node[above=2] {};
\draw[color=black,decorate,decoration={brace,raise=0.2cm},rotate=180]
(-4,0) -- (-2.1,0) node[below=8,pos=0.5] {2 r};
\draw[color=black,decorate,decoration={brace,raise=0.2cm},rotate=180]
(-1.9,0) -- (-1,0) node[below=8,pos=0.5] {r};
\end{tikzpicture}
\end{center}
\end{figure}

\noindent
with the GH center either in the middle or on the right depending on the solution considered.
\noindent
\begin{table}[H]
\caption{The 18 multicenter solutions with global D-brane charges $(Q_{D6},Q_{D4}^1,Q_{D4}^2,Q_{D4}^3;Q_{D2}^1,Q_{D2}^2,Q_{D2}^3,Q_{D0})=(2,0,0,0;1,1,1,0)$.}
\label{TableSol2}
\begin{center}
\begin{tabular}{|c|c||c|c|}
  \hline
  \hline
 	1 & \quad \parbox{0cm}{\begin{alignat}{2}
 		&\Gamma_0 &&\:=\:  (2,\,1,\,0,\,-2\:;\:0,\,1,\,0,\,0) \nonumber\\
 		&\Gamma_1 &&\:=\:  (0,\,-1,\,0,\,0\:;\:0,\,0,\,1,\,0)  \nonumber \\
 		&\Gamma_2 &&\:=\: (0,\,0,\,0,\,2\:;\:1,\,0,\,0,\,0) \nonumber
 	\end{alignat}}
	&
	   	2 &\quad \parbox{2cm}{\begin{alignat}{2}
		&\Gamma_0 &&\:=\:  (2,\,0,\,-1,\,0\:;\:0,\,0,\,0,\,0) \nonumber\\
		&\Gamma_1 &&\:=\:  (0,\,0,\,0,\,0\:;\:0,\,1,\,0,\,-1)  \nonumber \\
		&\Gamma_2 &&\:=\:  (0,\,0,\,1,\,0\:;\:1,\,0,\,1,\,1) \nonumber
		\end{alignat}}
 \\
  \hline
   	3 &\quad \parbox{2cm}{\begin{alignat}{2}
 		&\Gamma_0 &&\:=\:  (2,\,1,\,-2,\,0\:;\:0,\,0,\,1,\,0) \nonumber\\
 		&\Gamma_1 &&\:=\:  (0,\,0,\,2,\,0\:;\:1,\,0,\,0,\,0)  \nonumber \\
 		&\Gamma_2 &&\:=\: (0,\,-1,\,0,\,0\:;\:0,\,1,\,0,\,0)  \nonumber
 	\end{alignat}}
	&
	   	4 &\quad \parbox{2cm}{\begin{alignat}{2}
		&\Gamma_0 &&\:=\: (2,\,0,\,-2,\,1\:;\:1,\,0,\,0,\,0) \nonumber\\
		&\Gamma_1 &&\:=\:  (0,\,0,\,0,\,-1\:;\:0,\,1,\,0,\,0)  \nonumber \\
		&\Gamma_2 &&\:=\: (0,\,0,\,2,\,0\:;\:0,\,0,\,1,\,0) \nonumber
		\end{alignat}}
  \\
		\hline
   	5 &\quad \parbox{2cm}{\begin{alignat}{2}
 		&\Gamma_0 &&\:=\: (2,\,-2,\,0,1\:;\:0,\,1,\,0,\,0) \nonumber\\
 		&\Gamma_1 &&\:=\:  (0,\,2,\,0,\,0\:;\:0,\,0,\,1,\,0) \nonumber \\
 		&\Gamma_2 &&\:=\: (0,\,0,\,0,\,-1\:;\:1,\,0,\,0,\,0)  \nonumber
 	\end{alignat}}
	&
	   	6 &\quad \parbox{2cm}{\begin{alignat}{2}
		&\Gamma_0 &&\:=\: (2,\,0,\,0,\,-1\:;\:0,\,0,\,0,\,0) \nonumber\\
		&\Gamma_1 &&\:=\:  (0,\,0,\,0,\,1\:;\:1,\,1,\,0,\,1) \nonumber \\
		&\Gamma_2 &&\:=\: (0,\,0,\,0,\,0\:;\:0,\,0,\,1,\,-1)  \nonumber
		\end{alignat}}
  \\
  \hline
	  	7 &\quad \parbox{2cm}{\begin{alignat}{2}
		&\Gamma_0 &&\:=\: (2,\,0,\,1,\,-2\:;\:1,\,0,\,0,\,0) \nonumber\\
		&\Gamma_1 &&\:=\:  (0,\,0,\,0,\,2\:;\:0,\,1,\,0,\,0)  \nonumber \\
		&\Gamma_2 &&\:=\: (0,\,0,\,-1,\,0\:;\:0,\,0,\,1,\,0) \nonumber
		\end{alignat}}	
	&
	8 &\quad \parbox{2cm}{\begin{alignat}{2}
		&\Gamma_0 &&\:=\: (2,\,-1,\,0,\,0\:;\:0,\,0,\,0,\,0) \nonumber\\
		&\Gamma_1 &&\:=\: (0,\,1,\,0,\,0\:;\:0,\,1,\,1,\,1)  \nonumber \\
		&\Gamma_2 &&\:=\: (0,\,0,\,0,\,0\:;\:1,\,0,\,0,\,-1)  \nonumber
		\end{alignat}}	
	\\
	  \hline
9 &\quad \parbox{2cm}{\begin{alignat}{2}
	&\Gamma_0 &&\:=\: (2,\,-2,\,1,\,0\:;\:0,\,0,\,1,\,0) \nonumber\\
	&\Gamma_1 &&\:=\:  (0,\,0,\,-1,\,0\:;\:1,\,0,\,0,\,0), \nonumber \\
	&\Gamma_2 &&\:=\: (0,\,2,\,0,\,0\:;\:0,\,1,\,0,\,0)  \nonumber
	\end{alignat}}	
&
10 &\quad \parbox{2cm}{\begin{alignat}{2}
	&\Gamma_0 &&\:=\: (2,\,2,\,-1,\,0\:;\:0,\,0,\,1,\,0) \nonumber\\
	&\Gamma_1 &&\:=\:  (0,\,-2,\,0,\,0\:;\:0,\,1,\,0,\,0)  \nonumber \\
	&\Gamma_2 &&\:=\:  (0,\,0,\,1,\,0\:;\:1,\,0,\,0,\,0) \nonumber
	\end{alignat}}
\\	
	\hline
	11	& \quad \parbox{0cm}{\begin{alignat}{2}
		&\Gamma_0 &&\:=\:  (2,\,2,\,0,\,-1\:;\:0,\,1,\,0,\,0) \nonumber\\
		&\Gamma_1 &&\:=\:  (0,\,0,\,0,\,1\:;\:1,\,0,\,0,\,0)  \nonumber \\
		&\Gamma_2 &&\:=\: (0,\,-2,\,0,\,0\:;\:0,\,0,\,1,\,0) \nonumber
		\end{alignat}}
	&
	12 &\quad \parbox{2cm}{\begin{alignat}{2}
		&\Gamma_0 &&\:=\:  (2,\,0,\,-1,\,2\:;\:1,\,0,\,0,\,0) \nonumber\\
		&\Gamma_1 &&\:=\: (0,\,0,\,0,\,-2\:;\:0,\,1,\,0,\,0)  \nonumber \\
		&\Gamma_2 &&\:=\: (0,\,0,\,1,\,0\:;\:0,\,0,\,1,\,0)  \nonumber
		\end{alignat}}
	\\
	  \hline
  \end{tabular}
\end{center}
\end{table}
\begin{table} 
\begin{center}
\begin{tabular}{|c|c||c|c|}
  \hline
     	13 & \quad \parbox{0cm}{\begin{alignat}{2}
 		&\Gamma_0 &&\:=\: (2,\,0,\,2,\,-1\:;\:1,\,0,\,0,\,0)  \nonumber\\
 		&\Gamma_1 &&\:=\:  (0,\,0,\,-2,\,0\:;\:0,\,0,\,1,\,0)  \nonumber \\
 		&\Gamma_2 &&\:=\: (0,\,0,\,0,\,1\:;\:0,\,1,\,0,\,0), \nonumber
 	\end{alignat}}
	&       	14 &\quad \parbox{2cm}{\begin{alignat}{2}
		&\Gamma_0 &&\:=\: (2,\,-1,\,2,\,0\:;\:0,\,0,\,1,\,0)  \nonumber\\
		&\Gamma_1 &&\:=\: (0,\,0,\,-2,\,0\:;\:1,\,0,\,0,\,0)   \nonumber \\
		&\Gamma_2 &&\:=\: (0,\,1,\,0,\,0\:;\:0,\,1,\,0,\,0)  \nonumber
		\end{alignat}}
  \\
  \hline
  	       	15 &\quad \parbox{2cm}{\begin{alignat}{2}
  	&\Gamma_0 &&\:=\: (2,\,-1,\,0,\,2\:;\:0,\,1,\,0,\,0) \nonumber\\
  	&\Gamma_1 &&\:=\: (0,\,1,\,0,\,0\:;\:0,\,0,\,1,\,0)  \nonumber \\
  	&\Gamma_2 &&\:=\:  (0,\,0,\,0,\,-2\:;\:1,\,0,\,0,\,0) \nonumber
  	\end{alignat}}
	&
	     16 &\quad \parbox{2cm}{\begin{alignat}{2}
		&\Gamma_0 &&\:=\: (2,\,1,\,0,\,0\:;\:0,\,0,\,0,\,0) \nonumber\\
		&\Gamma_1 &&\:=\: (0,\,0,\,0,\,0\:;\:1,\,0,\,0,\,1)  \nonumber \\
		&\Gamma_2 &&\:=\: (0,\,-1,\,0,\,0\:;\:0,\,1,\,1,\,-1)  \nonumber
		\end{alignat}}	
  \\
  \hline
       	17 &\quad \parbox{2cm}{\begin{alignat}{2}
	&\Gamma_0 &&\:=\:  (2,\,0,\,0,\,1\:;\:0,\,0,\,0,\,0) \nonumber\\
	&\Gamma_1 &&\:=\:  (0,\,0,\,0,\,0\:;\:0,\,0,\,1,\,1) \nonumber \\
	&\Gamma_2 &&\:=\: (0,\,0,\,0,\,-1\:;\:1,\,1,\,0,\,-1)  \nonumber
	\end{alignat}}	
	&
       	18 &\quad \parbox{2cm}{\begin{alignat}{2}
 		&\Gamma_0 &&\:=\: (2,\,0,\,1,\,0\:;\:0,\,0,\,0,\,0) \nonumber\\
 		&\Gamma_1 &&\:=\: (0,\,0,\,0,\,0\:;\:0,\,1,\,0,\,1)  \nonumber \\
 		&\Gamma_2 &&\:=\: (0,\,0,\,-1,\,0\:;\:1,\,0,\,1,\,-1)  \nonumber
 	\end{alignat}}	
  \\
  \hline
\end{tabular}
\end{center}
\end{table}
\pagebreak
\section{Three-node abelian quiver with general (a,b,c)} \label{abcapp}
The case of a general Abelian 3-node quiver with closed loop and with vanishing FI parameters, is quite similar to $(a,b,c)=(2,1,1)$ . It is described by the following quiver
\begin{equation}
\begin{aligned}
\begin{tikzcd}
&\circled{1} \ar[dr,"Y_\beta"]& \\
\circled{1} \ar[ur,"X_\alpha"] & & \circled{1} \ar[ll,"C_\gamma"]  \\
\end{tikzcd} 
\end{aligned} \, ,
\end{equation}
with $\alpha=1, \dots , a, \,\,\beta= 1, \dots , b, \,\,\gamma = 1, \dots ,c$ where $(a,b,c)$ is the unique triplet of integer intersection product $(a,b,c)\equiv(\langle \Gamma_{i}, \Gamma_j \rangle,\langle \Gamma_{j}, \Gamma_k \rangle , \langle \Gamma_{k}, \Gamma_i \rangle )$, $i,j,k$ are three different integers between 0 and 2 in order to satisfy $a \geq b >0$ and $c>0$. The D-term equations are given by 
\begin{align}
\sum_{\alpha=1}^a |X_\alpha|^2 - \sum_{\gamma=1}^c|C_\gamma|^2 &= 0 \\
\sum_{\beta=1}^b |Y_\beta|^2 - \sum_{\alpha=1}^a |X_\alpha|^2 &= 0 \\
-\sum_{\beta=1}^b |Y_\beta|^2 + \sum_{\gamma=1}^c |C_\gamma|^2 &= 0 \, .
\end{align}
Again, we assume a generic cubic superpotential
\begin{align}
W &= w_{\alpha \beta \gamma} X_\alpha Y_\beta C_\gamma \, ,
\end{align}
which gives the following F-term equations:
\begin{align}
w_{\alpha \beta \gamma} Y_\beta C_\gamma &= 0, \qquad w_{\alpha \beta \gamma} X_\alpha C_\gamma =0,  \qquad  w_{\alpha \beta \gamma} X_\alpha Y_\beta=0  \, .
\end{align}
As argued in \cite{Bena:2012hf}, the solution space consists of 3 chambers, in each of which only one of the three fields vanishes. However by D-term equations, this also implies vanishing of all three fields. So there is only one chamber, consisting a single solution. Again, the solution preserves $U(1) \times U(1)$ gauge symmetry.
 
We briefly make comparison with  \cite{Bena:2012hf}, which considered same quiver, but with non-zero FI parameters, and came to rather different conclusions. For non-zero FI parameters, after setting one of the fields to zero, D-term equations define a product of projective spaces. On the other hand, setting a field to zero, solves two F-term equations automatically. The remaining one defines a complete intersection manifold in the product of projective spaces. 
Requiring the dimension of this manifold to be non-negative gives the condition $a +b \geq c +2$ and permutations. When we set the FI parameters to zero, these projective spaces collapse to a point and so does the intersection manifold. As a result we do not have any condition on $(a,b,c)$. This is rather puzzling as physically one would have expected to get some version of triangle inequality. In particular, we would like to understand the Coulomb branch description of quivers with  triangle inequality violating $(a,b,c)$. 

Due to the above mentioned differences, the conclusions of \cite{Bena:2012hf} do not apply to quivers discussed in this paper. 
\newpage

\bibliography{references}

\providecommand{\href}[2]{#2}\begingroup\raggedright\begin{thebibliography}{10}

\bibitem{Strominger:1996sh}
A.~Strominger and C.~Vafa,  {\em {Microscopic origin of the Bekenstein-Hawking
  entropy}}, Phys. Lett. {\bf B379} (1996) 99--104
[\href{http://www.arXiv.org/abs/hep-th/9601029}{{\tt hep-th/9601029}}].

\bibitem{David:2002wn}
J.~R. David, G.~Mandal and S.~R. Wadia,  {\em {Microscopic formulation of black
  holes in string theory}}, Phys. Rept. {\bf 369} (2002) 549--686
[\href{http://www.arXiv.org/abs/hep-th/0203048}{{\tt hep-th/0203048}}].

\bibitem{Sen:2007qy}
A.~Sen,  {\em {Black Hole Entropy Function, Attractors and Precision Counting
  of Microstates}}, Gen. Rel. Grav. {\bf 40} (2008) 2249--2431
[\href{http://www.arXiv.org/abs/0708.1270}{{\tt 0708.1270}}].

\bibitem{Dijkgraaf:1996xw}
R.~Dijkgraaf, G.~W. Moore, E.~P. Verlinde and H.~L. Verlinde,  {\em {Elliptic
  genera of symmetric products and second quantized strings}}, Commun. Math.
  Phys. {\bf 185} (1997) 197--209
[\href{http://www.arXiv.org/abs/hep-th/9608096}{{\tt hep-th/9608096}}].

\bibitem{Maldacena:1999bp}
J.~M. Maldacena, G.~W. Moore and A.~Strominger,  {\em {Counting BPS black holes
  in toroidal Type II string theory}},
\href{http://www.arXiv.org/abs/hep-th/9903163}{{\tt hep-th/9903163}}.

\bibitem{Shih:2005qf}
D.~Shih, A.~Strominger and X.~Yin,  {\em {Counting dyons in N=8 string
  theory}}, JHEP {\bf 06} (2006) 037
[\href{http://www.arXiv.org/abs/hep-th/0506151}{{\tt hep-th/0506151}}].

\bibitem{David:2006ud}
J.~R. David, D.~P. Jatkar and A.~Sen,  {\em {Dyon spectrum in generic N=4
  supersymmetric Z(N) orbifolds}}, JHEP {\bf 01} (2007) 016
[\href{http://www.arXiv.org/abs/hep-th/0609109}{{\tt hep-th/0609109}}].

\bibitem{Denef:2002ru}
F.~Denef,  {\em {Quantum quivers and Hall / hole halos}}, JHEP {\bf 0210}
  (2002) 023
[\href{http://www.arXiv.org/abs/hep-th/0206072}{{\tt hep-th/0206072}}].

\bibitem{Gaddam:2016xum}
N.~Gaddam,  {\em {Elliptic genera from multi-centers}}, JHEP {\bf 05} (2016)
  076
[\href{http://www.arXiv.org/abs/1603.01724}{{\tt 1603.01724}}].

\bibitem{Chowdhury:2014yca}
A.~Chowdhury, R.~S. Garavuso, S.~Mondal and A.~Sen,  {\em {BPS State Counting
  in N=8 Supersymmetric String Theory for Pure D-brane Configurations}}, JHEP
  {\bf 10} (2014) 186
[\href{http://www.arXiv.org/abs/1405.0412}{{\tt 1405.0412}}].

\bibitem{Chowdhury:2015gbk}
A.~Chowdhury, R.~S. Garavuso, S.~Mondal and A.~Sen,  {\em {Do All BPS Black
  Hole Microstates Carry Zero Angular Momentum?}}, JHEP {\bf 04} (2016) 082
[\href{http://www.arXiv.org/abs/1511.06978}{{\tt 1511.06978}}].

\bibitem{Denef:2000nb}
F.~Denef,  {\em {Supergravity flows and D-brane stability}}, JHEP {\bf 08}
  (2000) 050
[\href{http://www.arXiv.org/abs/hep-th/0005049}{{\tt hep-th/0005049}}].

\bibitem{Denef:2001xn}
F.~Denef, B.~R. Greene and M.~Raugas,  {\em {Split attractor flows and the
  spectrum of BPS D-branes on the quintic}}, JHEP {\bf 05} (2001) 012
[\href{http://www.arXiv.org/abs/hep-th/0101135}{{\tt hep-th/0101135}}].

\bibitem{Denef:2001ix}
F.~Denef,  {\em {(Dis)assembling special Lagrangians}},
\href{http://www.arXiv.org/abs/hep-th/0107152}{{\tt hep-th/0107152}}.

\bibitem{Bates:2003vx}
B.~Bates and F.~Denef,  {\em {Exact solutions for supersymmetric stationary
  black hole composites}}, JHEP {\bf 1111} (2011) 127
[\href{http://www.arXiv.org/abs/hep-th/0304094}{{\tt hep-th/0304094}}].

\bibitem{Denef:2007vg}
F.~Denef and G.~W. Moore,  {\em {Split states, entropy enigmas, holes and
  halos}}, JHEP {\bf 11} (2011) 129
[\href{http://www.arXiv.org/abs/hep-th/0702146}{{\tt hep-th/0702146}}].

\bibitem{Bena:2013dka}
I.~Bena and N.~P. Warner,  {\em {Resolving the Structure of Black Holes:
  Philosophizing with a Hammer}},
\href{http://www.arXiv.org/abs/1311.4538}{{\tt 1311.4538}}.

\bibitem{Sen:2005wa}
A.~Sen,  {\em {Black hole entropy function and the attractor mechanism in
  higher derivative gravity}}, JHEP {\bf 09} (2005) 038
[\href{http://www.arXiv.org/abs/hep-th/0506177}{{\tt hep-th/0506177}}].

\bibitem{Sen:2008vm}
A.~Sen,  {\em {Quantum Entropy Function from AdS(2)/CFT(1) Correspondence}},
  Int. J. Mod. Phys. {\bf A24} (2009) 4225--4244
[\href{http://www.arXiv.org/abs/0809.3304}{{\tt 0809.3304}}].

\bibitem{Bena:2018bbd}
I.~Bena, P.~Heidmann and D.~Turton,  {\em {AdS$_2$ Holography: Mind the Cap}},
\href{http://www.arXiv.org/abs/1806.02834}{{\tt 1806.02834}}.

\bibitem{Avila:2017pwi}
J.~Avila, P.~F. Ramirez and A.~Ruiperez,  {\em {One Thousand and One Bubbles}},
\href{http://www.arXiv.org/abs/1709.03985}{{\tt 1709.03985}}.

\bibitem{Balasubramanian:2006gi}
V.~Balasubramanian, E.~G. Gimon and T.~S. Levi,  {\em {Four Dimensional Black
  Hole Microstates: From D-branes to Spacetime Foam}}, JHEP {\bf 01} (2008) 056
[\href{http://www.arXiv.org/abs/hep-th/0606118}{{\tt hep-th/0606118}}].

\bibitem{Bena:2015bea}
I.~Bena, S.~Giusto, R.~Russo, M.~Shigemori and N.~P. Warner,  {\em {Habemus
  Superstratum! A constructive proof of the existence of superstrata}}, JHEP
  {\bf 05} (2015) 110
[\href{http://www.arXiv.org/abs/1503.01463}{{\tt 1503.01463}}].

\bibitem{Mateos:2001qs}
D.~Mateos and P.~K. Townsend,  {\em {Supertubes}}, Phys. Rev. Lett. {\bf 87}
  (2001) 011602
[\href{http://www.arXiv.org/abs/hep-th/0103030}{{\tt hep-th/0103030}}].

\bibitem{Bena:2008wt}
I.~Bena, N.~Bobev and N.~P. Warner,  {\em {Spectral Flow, and the Spectrum of
  Multi-Center Solutions}}, Phys. Rev. {\bf D77} (2008) 125025
[\href{http://www.arXiv.org/abs/0803.1203}{{\tt 0803.1203}}].

\bibitem{Bena:2009en}
I.~Bena, S.~Giusto, C.~Ruef and N.~P. Warner,  {\em {Multi-Center non-BPS Black
  Holes: the Solution}}, JHEP {\bf 11} (2009) 032
[\href{http://www.arXiv.org/abs/0908.2121}{{\tt 0908.2121}}].

\bibitem{Vasilakis:2011ki}
O.~Vasilakis and N.~P. Warner,  {\em {Mind the Gap: Supersymmetry Breaking in
  Scaling, Microstate Geometries}}, JHEP {\bf 10} (2011) 006
[\href{http://www.arXiv.org/abs/1104.2641}{{\tt 1104.2641}}].

\bibitem{Heidmann:2017cxt}
P.~Heidmann,  {\em {Four-center bubbled BPS solutions with a Gibbons-Hawking
  base}},
\href{http://www.arXiv.org/abs/1703.10095}{{\tt 1703.10095}}.

\bibitem{Bena:2011zw}
I.~Bena, B.~D. Chowdhury, J.~de~Boer, S.~El-Showk and M.~Shigemori,  {\em
  {Moulting Black Holes}}, JHEP {\bf 1203} (2012) 094
[\href{http://www.arXiv.org/abs/1108.0411}{{\tt 1108.0411}}].

\bibitem{Heidmann:2017toappear}
P.~Heidmann,  {\em {To appear}},.

\bibitem{Bachas:1996bp}
C.~Bachas and E.~Kiritsis,  {\em {$F^4$} terms in {$N=4$} string vacua}, Nucl.
  Phys. Proc. Suppl. {\bf 55B} (1997) 194
  [\href{http://www.arXiv.org/abs/hep-th/9611205}{{\tt hep-th/9611205}}].

\bibitem{Gregori:1997hi}
A.~Gregori, E.~Kiritsis, C.~Kounnas, N.~A. Obers, P.~M. Petropoulos and
  B.~Pioline,  {\em {$R^2$} corrections and nonperturbative dualities of {$N =
  4$} string ground states}, Nucl. Phys. {\bf B510} (1998) 423
  [\href{http://www.arXiv.org/abs/hep-th/9708062}{{\tt hep-th/9708062}}].

\bibitem{Bianchi:2017bxl}
M.~Bianchi, J.~F. Morales, L.~Pieri and N.~Zinnato,  {\em {More on microstate
  geometries of 4d black holes}}, JHEP {\bf 05} (2017) 147
[\href{http://www.arXiv.org/abs/1701.05520}{{\tt 1701.05520}}].

\bibitem{Manschot:2010qz}
J.~Manschot, B.~Pioline and A.~Sen,  {\em {Wall Crossing from Boltzmann Black
  Hole Halos}}, JHEP {\bf 07} (2011) 059
[\href{http://www.arXiv.org/abs/1011.1258}{{\tt 1011.1258}}].

\bibitem{Sen:2009vz}
A.~Sen,  {\em {Arithmetic of Quantum Entropy Function}}, JHEP {\bf 08} (2009)
  068
[\href{http://www.arXiv.org/abs/0903.1477}{{\tt 0903.1477}}].

\bibitem{Dabholkar:2010rm}
A.~Dabholkar, J.~Gomes, S.~Murthy and A.~Sen,  {\em {Supersymmetric Index from
  Black Hole Entropy}}, JHEP {\bf 04} (2011) 034
[\href{http://www.arXiv.org/abs/1009.3226}{{\tt 1009.3226}}].

\bibitem{Douglas:1996sw}
M.~R. Douglas and G.~W. Moore,  {\em {D-branes, quivers, and ALE instantons}},
\href{http://www.arXiv.org/abs/hep-th/9603167}{{\tt hep-th/9603167}}.

\bibitem{Bena:2012hf}
I.~Bena, M.~Berkooz, J.~de~Boer, S.~El-Showk and D.~Van~den Bleeken,  {\em
  {Scaling BPS Solutions and pure-Higgs States}}, JHEP {\bf 1211} (2012) 171
[\href{http://www.arXiv.org/abs/1205.5023}{{\tt 1205.5023}}].

\bibitem{Dabholkar:2012nd}
A.~Dabholkar, S.~Murthy and D.~Zagier,  {\em {Quantum Black Holes, Wall
  Crossing, and Mock Modular Forms}},
\href{http://www.arXiv.org/abs/1208.4074}{{\tt 1208.4074}}.

\bibitem{Gauntlett:2002nw}
J.~P. Gauntlett, J.~B. Gutowski, C.~M. Hull, S.~Pakis and H.~S. Reall,  {\em
  {All supersymmetric solutions of minimal supergravity in five- dimensions}},
  Class.Quant.Grav. {\bf 20} (2003) 4587--4634
[\href{http://www.arXiv.org/abs/hep-th/0209114}{{\tt hep-th/0209114}}].

\bibitem{Bena:2013ora}
I.~Bena, S.~F. Ross and N.~P. Warner,  {\em {On the Oscillation of Species}},
  JHEP {\bf 1409} (2014) 113
[\href{http://www.arXiv.org/abs/1312.3635}{{\tt 1312.3635}}].

\bibitem{Bena:2010gg}
I.~Bena, N.~Bobev, S.~Giusto, C.~Ruef and N.~P. Warner,  {\em {An
  Infinite-Dimensional Family of Black-Hole Microstate Geometries}}, JHEP {\bf
  03} (2011) 022 [\href{http://www.arXiv.org/abs/1006.3497}{{\tt 1006.3497}}],
[Erratum: JHEP04,059(2011)].

\bibitem{Bena:2016ypk}
I.~Bena, S.~Giusto, E.~J. Martinec, R.~Russo, M.~Shigemori, D.~Turton and N.~P.
  Warner,  {\em {Smooth horizonless geometries deep inside the black-hole
  regime}}, Phys. Rev. Lett. {\bf 117} (2016), no.~20, 201601
[\href{http://www.arXiv.org/abs/1607.03908}{{\tt 1607.03908}}].

\bibitem{DallAgata:2010srl}
G.~Dall'Agata, S.~Giusto and C.~Ruef,  {\em {U-duality and non-BPS solutions}},
  JHEP {\bf 02} (2011) 074
[\href{http://www.arXiv.org/abs/1012.4803}{{\tt 1012.4803}}].

\end{thebibliography}\endgroup
\bibliographystyle{utphysmodb}

\end{document}